\documentclass[preprint]{aastex}

\slugcomment{} \shorttitle{EL-ICA of galaxy spectra}
\shortauthors{Lu et al.}

\begin{document}

\title{Ensemble Learning Independent Component Analysis of Normal Galaxy Spectra}
\author{Honglin Lu\altaffilmark{1}, Hongyan Zhou\altaffilmark{1}, Junxian Wang\altaffilmark{1},
Tinggui Wang\altaffilmark{1}, \\
Xiaobo Dong\altaffilmark{1}, Zhenquan Zhuang\altaffilmark{2}, and
Cheng Li\altaffilmark{1}}

\altaffiltext{1}{Center for Astrophysics, University of Science
and Technology of China, Hefei, Anhui, P.R.China}

\altaffiltext{2}{Department for Electronics of Science \&
technology, University of Science and Technology of China, Hefei,
Anhui, P.R.China}

\email{mtzhou@ustc.edu.cn}

\begin{abstract}

In this paper, we employe a new statistical analysis technique,
Ensemble Learning for Independent Component Analysis (EL-ICA), on
the synthetic galaxy spectra from a newly released high resolution
evolutionary model by Bruzual \& Charlot. We find that EL-ICA can
sufficiently compress the synthetic galaxy spectral library to 6
non-negative Independent Components (ICs), which are good
templates to model huge amount of normal galaxy spectra, such as
the galaxy spectra in the Sloan Digital Sky Survey (SDSS).
Important spectral parameters, such as starlight reddening,
stellar velocity dispersion, stellar mass and star formation
histories, can be given simultaneously by the fit. Extensive tests
show that the fit and the derived parameters are reliable for
galaxy spectra with the typical quality of the SDSS.

\end{abstract}


\section{Introduction}
With the coming of large sky area spectroscopic surveys, the first
decade of the $21^{st}$ century turns into the golden age for
extragalactic astronomy. The ongoing SDSS (the Sloan Digital Sky
Survey, York et al. 2000) and the forthcoming LAMOST (the Large
Sky Area Multi-Object Fiber Spectroscopic Telescope, Chu \& Zhao
1998) are now and will be providing us millions of high quality
optical spectra for both normal and active galaxies. For the first
time it becomes doable to explore the spectral properties of a
great diversity of galaxy populations. Strong constraints can be
set on various scenarios of galaxy formation and evolution by
comparing the measured quantities with theoretical predictions.
However, full exploitation of such large data sets demands for the
development of new methods that can extract the spectral
characteristics of individual galaxies reliably, consistently, and
efficiently.

In optical band, an observed galaxy spectrum is the combination of
starlight, nebular emission lines, and/or nuclear emission. Proper
interpretation of the stellar spectrum is essential to study the
galactic properties, such as dust contents, stellar velocity
dispersion, metallicity, and star formation histories. Nebular
emission lines are produced in the HII regions around young,
massive (OB) UV-emitting stars, and are important tools for
studying star formation and chemical evolution in galaxies. They
are often heavily entangled with stellar absorption features,
especially in weak emission line galaxies. Hence the proper
decomposition of emission lines and starlight components is
required toward a physical interpretation of a galaxy spectrum
(e.g., Tremonti et al. 2004). Proper stellar subtraction is also
essential for the study of AGN emission lines if contaminated by
the stellar absorption lines (e.g., Hao et al. 2002).

Many attempts have been made to fit the galaxy spectra in the past
decades. For those who are only interested in emission line
properties, the spectra of absorption line galaxies were generally
used as templates to fit the emission line free region of galaxy
spectra (e.g., Filippenko \& Sargent 1988). The ``pure line
spectrum'' can then be obtained by subtracting the best fitting
spectrum. The drawback of such an approach is obvious, since the
stellar content of an emission line galaxy under study is often
quite different from that of an absorption line galaxy.
Improvements have been made along this direction ever since. For
instance, Ho, Filippenko, \& Sargent (1997) used an objective
algorithm to find the best combination of galaxy spectra to create
an ``effective'' template. The use of a large basis of input
spectra ensures a closer match to the true stellar population.
Recently, Hao et al. (2005) applied Principle Component Analysis
(PCA, Madgwick et al. 2002 and references therein) to several
hundred pure absorption line galaxies and chose the first 8
eigen-spectra as the templates. This approach can greatly reduce
the computational time of the fit, and is ideal for fitting large
data sets like the SDSS.

To study the stellar contents reside in unresolved galaxies
requires the so called stellar population analysis. Two different
approaches have been adopted for this purpose: empirical (e.g.
Pelat 1997; Cid Fernandes et al. 2001) and evolutionary population
synthesis (pioneered by Tinsley 1967). The empirical population
synthesis uses the observed spectra of stars and star clusters to
model a galaxy spectrum. However, this method is limited by the
coverage range of the observed stellar library. The evolutionary
population synthesis is a more direct approach in theoretical
aspect. In this scheme, the age- and metallicity-dependent
modelled spectra of stellar systems are built with physical input
parameters, such as the Initial Mass Function (IMF), the Star
Formation Rate (SFR) and, in some cases, the rate of chemical
enrichment. The limitations of evolutionary population synthesis
models used to arise from the uncertainties in its two principal
building blocks: stellar evolution models and spectral libraries.
Great advances have been achieved in both aspects in recent years
(Charbonnel et al. 1999; Girardi et al. 2000; Le Borgne et al.
2003). The most up-to-date release of a high spectral resolution
stellar evolutionary synthesis model by Bruzual \& Charlot (2003,
hereafter BC03), which incorporates these progresses, provides a
library of Simple Stellar Populations (SSPs\footnote{SSPs are
defined as stellar populations whose star formation duration is
short in comparison with the lifetime of their most massive stars.
}). This spectral library enables accurate studies of stellar
populations in galaxies for the first time.

The BC03 library has been employed by several groups to decompose
galaxy spectra into various number of SSPs of different ages and
metallicities. To reduce the computational expense, either a
compressed version of the spectrum in study is used (BC03), or a
few SSPs are selected as templates (Tremonti et al. 2004). We note
that recent improvement in the statistical analysis technique may
help. In this work we adopt a rather different approach grounded
on Independent Component Analysis (ICA) to model galaxy spectra.
The most prominent earmark of this method in comparison with
previous ones is that it works without any a-priori knowledge on
the galaxy spectrum analyzed.

The first astrophysical application of ICA is the Cosmic Microwave
Background analysis (Baccigalupi et al. 2000) and it has
subsequently been implemented in this research area by several
authors (e.g. Maino et al. 2002; Moudden et al. 2004). We present
here the application of a new ICA algorithm---Ensemble Learning
for ICA (EL-ICA )---first proposed by Lappalainen (1998) and later
on generalized by Miskin \& MacKay (2001), to the spectral library
of synthetic galaxy spectra from BC03 and the spectral features of
the SDSS galaxies. The robustness and high efficiency of the new
method make it a viable technique for enormous data sets of normal
galaxy spectra, including type 2 AGN (Active Galactic Nuclei). Its
application to decompose stellar and nuclear components in the
spectrum of type 1 AGN will be explored in forthcoming papers.

This paper is organized as follows. In \S2 below, ICA method is
briefly introduced and EL algorithm is explained. \S3 applies the
EL-ICA method to the synthetic galaxy spectra from the
evolutionary models of BC03 SSP and derives our galaxy templates.
In \S4, we implement these templates to analyze the galaxy spectra
in the SDSS Data Release 2 (DR2). The reliability of the obtained
spectral parameters is carefully inspected in \S5. Our conclusions
are summarized in the last section. Throughout this paper, a
$\Lambda$-dominated cosmology with $H_0=70~km~ s^{-1}~Mpc^{-1}$,
$\Omega_M=$ 0.3 and $\Omega_\Lambda=$ 0.7 is assumed.

\section{Ensemble Learning Independent Component Analysis}
Here we only recall the fundamental concepts of ICA and the
essential features of EL algorithm. Readers who are interested in
ICA and EL method may refer to Hyv\"{a}rinen et al. (2001) and
Miskin \& MacKay (2001).

\subsection{Independent Component Analysis}

First introduced in the early 1980s and widely used in the late
1990s, ICA is a new Blind Source Separation (BSS) method for
finding statistically independent underlying components from
multidimensional statistical data. ICA can be defined as follows:
Given a set of observations of random variables
{$f^{j}(\lambda),~j=1,2,...,m$}, where $\lambda$ is sample index
(here we take $f^{j}(\lambda) $ as the spectrum of a stellar
system and $\lambda$ the wavelength), we can assume that they are
a linear combination of a number of Independent Components (ICs),
\begin{equation}
\label{eq1}
f^{j}(\lambda)=\sum_{i=1}^{n}a^{j}_{~i}IC^{i}(\lambda)+N^{j}(\lambda)
\end{equation}
where $a^{j}_{~i}$ is $m\times n$ mixing matrix of unknown
coefficients and $N^{j}(\lambda)$ is the noise of
$f^{j}(\lambda)$. The task of ICA is to estimate both the unknown
combining coefficients $a^{j}_{~i}$, the noise $N^{j}(\lambda)$
and the $IC^{i}(\lambda)$ from $f^{j}(\lambda)$ only based on the
statistical independence among $IC^{i}(\lambda)$.

As a useful statistical and computational technique, ICA can be
taken as an extension to Principle Component Analysis (PCA), but
much more powerful. In many cases, ICA can find the underlying
factors when PCA fails completely. The reason lies in two aspects:
1) Independence is much more stronger than uncorrelatedness. PCA
always finds eigen-vectors (or PCs) grounded on the fact that they
are uncorrelated, but uncorrelatedness is often not enough. 2)
Orthogonality is invalid for a general mixing process. ICA model
does not require the underlying factors are orthogonal while PCA
does. Nonlinear de-correlation is the basic method to find ICs.
Many methods have been developed for estimating the ICA model and
all of these methods use some form of high-order statistics.

\subsection{Ensemble Learning for ICA}

It has been shown that the EL algorithm can be applied to ICA
model (Lappalainen 1998). In information theory, entropy is
measure of uncertainty and it can be used as a measure of the
quality of our models. Suppose we don't know the distribution,
$p(\Theta)$, and our model is $q(\Theta)$, the relative entropy,
$d_{KL}(q||p)$, is a measure of how different the two probability
distributions (over the same event space) are. We can minimize
$d_{KL}(q||p)$ to have a probabilistic model as accurate as
possible. The EL algorithm is a low-cost method to approximate an
intractable posterior distribution, p, by a tractable separable
distribution, q. The optimum ensemble distributions have simple
forms if conjugate priors for the distributions are chosen (Miskin
\& MacKay 2001).

By adding a cost associated with increasing model complexity, the
EL algorithm can select the simplest interpretation of the data so
that the chance of over-fitting of the data is greatly reduced.
Assigning some specific form of prior to the ICs, the mixing
matrix and the noise, the EL algorithm can be applied to the ICA
model. The EL ICA model can further be extended to include the
notion of positivity of both the combination coefficients
$a^{i}_{~j}$ and the hidden sources $IC^{j}(\lambda)$ (Miskin \&
MacKay 2001). Therefore this algorithm is especially adapted to
our present purpose to derive minimizing number of non-negative
galaxy templates but representative to include information in the
galaxy spectra as much as possible. In this work, we assign an
exponential prior (which is the simplest one) to the ICs, a
rectified Gaussian\footnote{The rectified Gaussian distribution,
$\mathcal{G}^{rect}(\Theta)$, is equivalent to the Gaussian
distribution, $\mathcal{G}(\Theta)$, but zero for $\Theta<0$.}
prior to the mixing matrix, and a Gaussian noise model.

\section{EL-ICA of Synthetic Galaxy Spectra}

Using evolutionary stellar population synthesis model, BC03
generated a high resolution library of synthetic galaxy spectra,
GALAXEV \footnote{The GALAXEV package, including the library of
template stellar population spectra, the codes that used to
compute the library, and the corresponding references, can be
downloaded from http://www.cida.ve/~bruzual/bc2003.}, based on a
new observed spectral library of stars, STELIB (Le Borgne et al.
2003). This library spans wide range of ages between $1\times10^5$
and $2\times10^{10}$ yr. The model spectra were calculated for 6
different metallicities (0.005, 0.02, 0.2, 0.4, 1.0 and 2.5 times
of $Z\odot$) at a resolution of 3 \AA~FWHM across the wavelength
range of $3200~-~9500$ \AA~(corresponding to a median resolving
power $\lambda/\Delta\lambda\approx2000$). The total number of the
model spectra is 1326, 221 for each of the 6 metallicities (see
Fig. 1 for the spectra distribution in the age/metallicity plane).
Synthetic spectra with a larger wavelength range of
$91~\AA~-160~\mu m$, at lower resolution are also computed.

\subsection{EL-ICA of the SSP Spectra}

\begin{figure}
\epsscale{0.8} \plotone{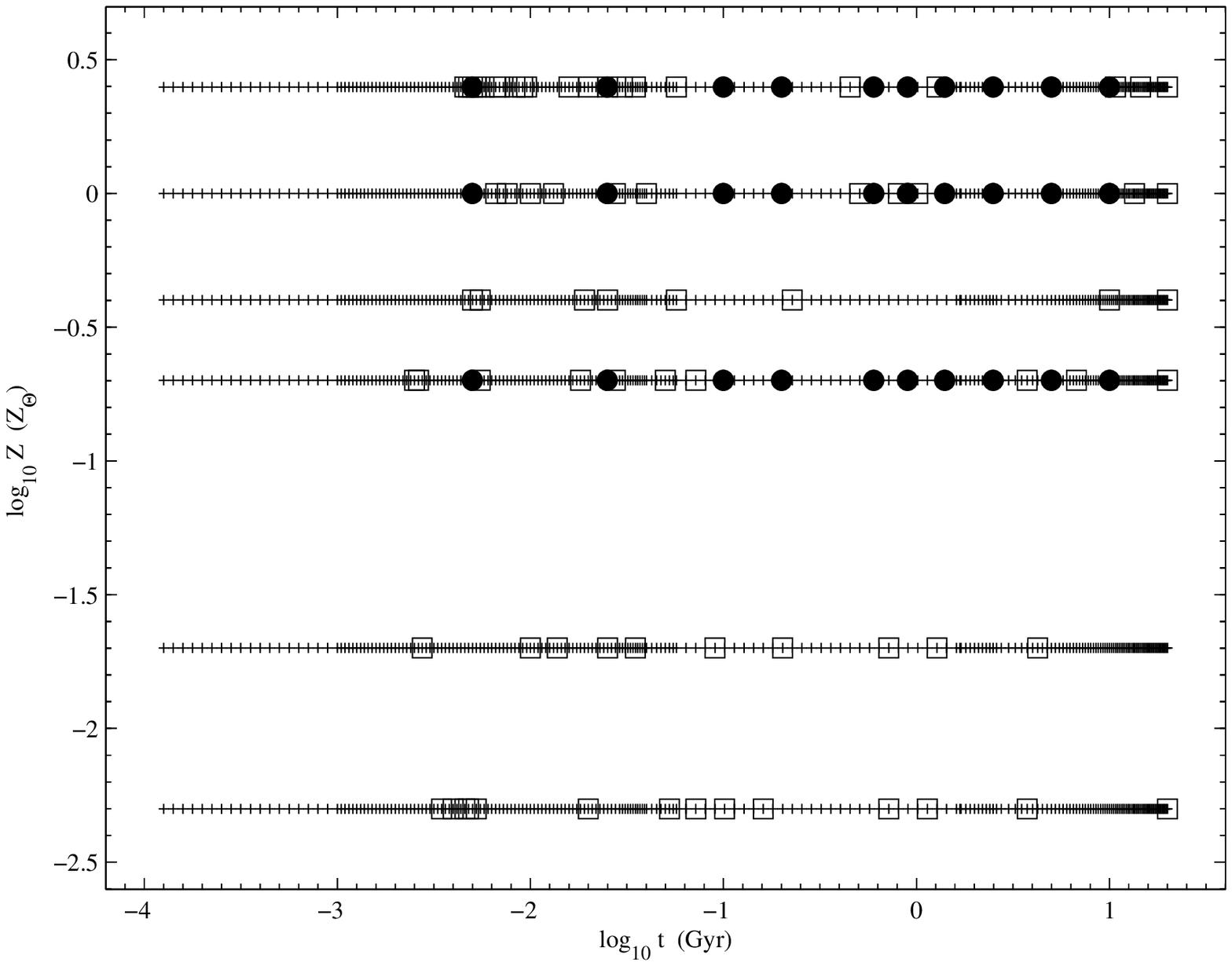} \caption{Distribution of the
blindly-selected SSP spectra (denoted as ``$\Box$'') employed for
Ensemble Learning Component Independent Analysis in the age vs.
metallicity space. The entire SSP spectra in the BC03 library
(denoted as ``+'') and the spectra used as templates by Tremonti
et al. (2004, denoted as ``$\bullet$'') are also shown for
comparison. } \label{f1}
\end{figure}

In this subsection we apply the EL-ICA method to the high
resolution SSP spectra in the GALAXEV. The 1326 SSP spectra are
first truncated to wavelength range of $3322-9200$ \AA~ (the short
wavelength limit is set by the high resolution wavelength range of
the GALAXEV and the long wavelength limit is chosen to match that
of the SDSS spectrograph) before further analyses. As a high-order
statistical algorithm, EL-ICA converges very slow. In this paper,
we only apply EL-ICA on a small subsample of the SSP spectra. 74
out of the 1326 SSP spectra are picked up as a subsample following
the procedures below. First we select the eldest SSP spectrum (age
= $2\times10^{10}$ yr) with the highest metallicity ($5~Z\odot$,
the upper right one in Figure \ref{f1}) as a start point. In step
2 we move to the second highest metallicity ( $Z=Z\odot$): the 221
spectra are examined in the order of decreasing age, and the first
one which is significantly different from the spectrum/spectra in
the subsample is added to the subsample. The definition of
``significantly different'' is
\begin{equation}\label{eq2}
\sqrt{\frac{1}{C}\sum_{k=1}^{C}\{\frac{[SSP^i(\lambda_k)-SSP^j(\lambda_k)]}{[SSP^i(\lambda_k)+SSP^j(\lambda_k)]/2}\}^2}~>~\frac{3}{50},
\end{equation}
where C is the number of wavelength channels, and $\frac{3}{50}$
stands for 3 $\sigma $ significance level for the SSP spectra with
standard S/N of $\sim 50$. Then we move to the next lower
metallicity, and repeat the above selection. Once we reach the
lowest metallicity, we turn over to the highest metallicity one,
repeat the above steps, and so on. The subsample is selected to
statistically represent the whole sample on the standard S/N level
of $\sim 50$. The distribution of the derived subsample is
displayed in Figure \ref{f1}, where we can see that the 74 SSP
spectra cover a wide range of age and metallicity. Note that there
is no SSP younger than $\sim 2.5\times 10^{6}$ yr in our
subsample. This is actually expected since that at $age\lesssim
10^6$ yr, the SSP are dominated by the hottest massive stars and
the spectra are quite similar across the wavelength range of
$3322-9200$ \AA. We also plot in Figure \ref{f1} the
age/metallicity coverage of a sample of stellar populations used
by Tremonti et al. (2004). The main difference between the two is
that our sample includes 6 metallicities while Tremonti et al.'s
includes 3, and our sample is selected unsupervised.

\begin{figure}
\epsscale{0.8} \plotone{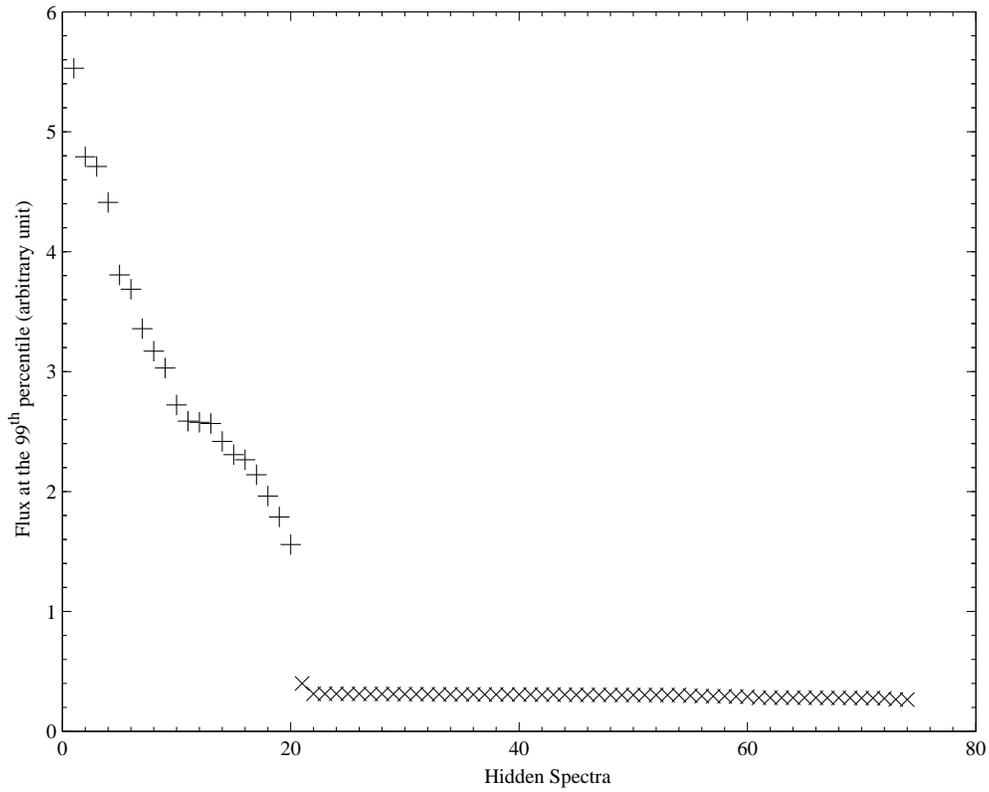} \caption{Distribution of the
flux at the $99^{th}$ percentile of the 74 recovered Hidden
Spectra. It is remarkable that flux of the first 20 Hidden Spectra
(denoted as ``+'') is much larger than that of the last 54
(denoted as ``$\times$''s). The Hidden Spectra are sorted as
descending of flux for visualization. } \label{f2}
\end{figure}

\begin{figure}
\epsscale{0.8} \plotone{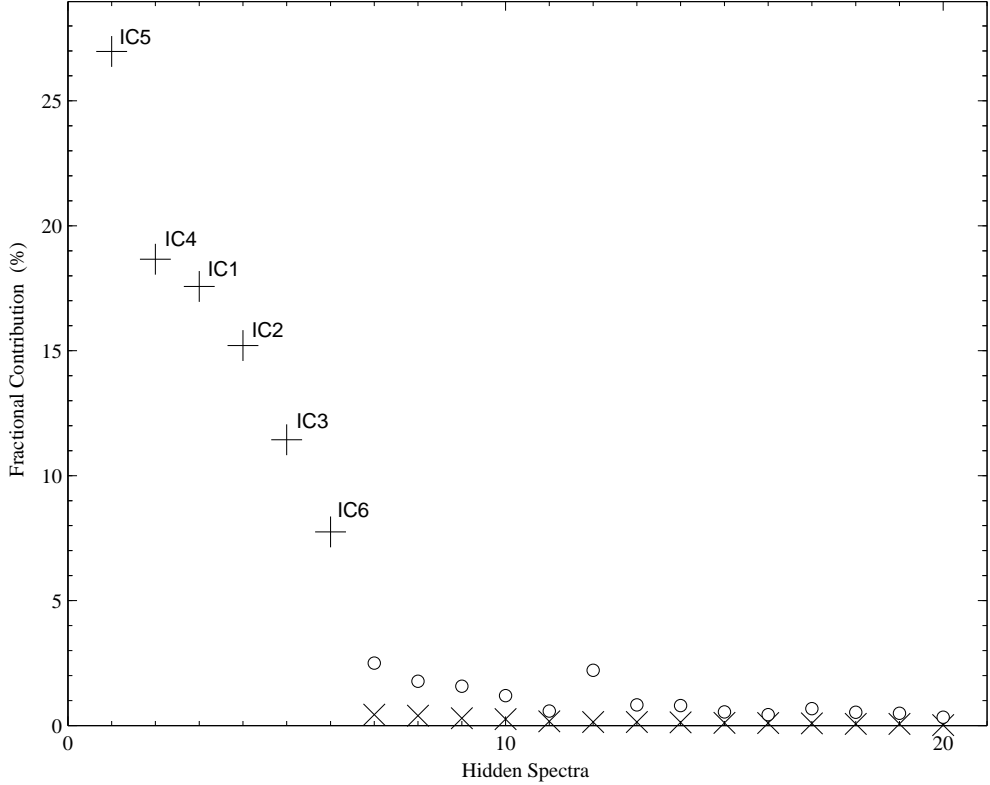} \caption{Distribution of the
average fractional contribution of the 20 Hidden Spectra selected
in Figure \ref{f2} to the 1,326 SSP spectra in the BC03 library.
The Hidden Spectra are sorted as descending fractional
contribution for clarity. It can be seen that the contribution of
the first 6 Hidden Spectra (denoted as ``+'') is significantly
larger than that of the last 14 (denoted as ``$\times $''). Even
the largest fractional contribution of the last 14 Hidden Spectra
to the individual SSP spectrum (denoted as ``$\circ$'') is
$<2.5\%$, and their cumulative contribution is $\simeq 2.4\% $.
The 6 ICs are named in the order of their ``spectral types'' (see
the text and Figure \ref{f4}). } \label{f3}
\end{figure}

The EL-ICA method is applied to the 74 SSP spectra selected above,
and 74 Hidden Spectra (HS\footnote{We denote all the spectra
obtained by EL-ICA as ``HS'' (Hidden Spectra), and the spectra
finally found to be informative as ``ICs'' (Independent
Components). }) are derived. In Figure \ref{f2} we plot the flux
at the $99^{th}$ percentile level for each of the 74 hidden
spectrum. It is obvious that 54 of the hidden spectra have almost
zero flux, comparing with the other 20. Then the first 20 HS are
normalized at 5500 \AA~ and utilized to fit the 1,326 BC03 SSP
spectra through linear least-squares fitting with non-negativity
constraints,
\begin{equation}\label{eq3}
SSP^{j}(\lambda) = \sum_{i=1}^{20}
c^{j}_{~i}~HS^{i}(\lambda),~~c^{j}_{~i}\geq 0,~j=1,2,...,1326.
\end{equation}
We define the fractional contribution of the $i^{th}$ HS to the
$j^{th}$ SSP as the normalized best fitting coefficient,
$c^j_{~i,norm}=\frac{c^j_{~i}}{\sum_{i=1}^{20}c^j_{~i}}\times
100\%$, and the average fractional contribution of the $i^{th}$ HS
to the BC03 SSP,
$c_{i}=\frac{\sum_{j=1}^{1326}c^j_{~i,norm}}{1326}$. The
distribution of the average fractional contribution of the 20 HS
is shown in Figure \ref{f3}. It can be seen that the contribution
of the first 6 HS is significantly larger than that of the rest
ones. The accumulative contribution of the first 6 HS amounts to
$\sim 97.6\%$, and that of the last 14 HS is only $\sim 2.4\%$. We
obtain similar results while fitting the spectra of the SDSS
galaxies with the 20 HS (see \S4). The 6 hidden spectra are
therefor chosen as the final Independent Components (ICs), the
spectra of which are presented in Figure \ref{f4}. Actually none
of the contribution of the last 14 HS to the individual SSP
spectrum is larger than $2.5\%$.

\subsection{Interpretation of the IC Spectra}
Though EL-ICA, we compressed the 1326 BC03 SSP spectra into 6
non-negative ICs. The physical meaning of the 6 ICs is interesting
and could be easily understood through examining the spectra
visually. The 6 ICs are named in the order of the spectral type
(``early---late'').
\begin{description}
  \item[IC1]
  represents the blue continuum of O star. However, the CaII H
  and K lines at 3933 and 3968 \AA, which are stronger in late-type
  stars than in early-type stars, also exist;
  \item[IC2]
  is similar to the spectrum of B star. Nevertheless, the absorption lines of
  neutral metals and molecules such as TiO and titanium Oxide,
  which are normal in spectra of later than K stars, are also
  identified;
  \item[IC3]
  shows extremely strong Balmer absorption lines and Balmer jump
  in the blue part of the spectra, even more prominent than A
  stars. But the Ca II triplet at 8498, 8542, and 8662 \AA,
  and more lines of neutral metals and molecules appear in the red;
  \item[IC4~ and IC5]
  are somewhat like hybrids of F to K stars, with stronger neutral metal and
  molecule lines.
  \item[IC6]
  is similar to the spectrum of M star in the long wavelength range but shows high
  order Balmer absorption lines in the short wavelength.
\end{description}

The spectral properties of the ICs imply that a tight correlation
between the stellar population age and the ICs. In Figure \ref{f5}
we plot the contribution of 6 ICs to individual SSP spectrum as a
function of age for 6 metallicities, where clear correlations
between the contributions of ICs and the age can be seen. For
instance, the fractional contribution of $IC1\gtrsim 8\%$
indicates an age of $t\lesssim 10^8$ yr, and $IC3\gtrsim 30\%$
indicates $10^8\lesssim t\lesssim10^9 $, corresponding the age of
the so-called ``E+A'' galaxies (e.g., Dressler \& Gunn 1983).

According the EL-ICA model, a spectrum of a stellar system
corresponds to a vector in the 6-D IC space. In Figure \ref{f6} we
plot the projection of the SSP spectra on the IC versus IC planes
to give a visual impression of the distribution of stellar systems
in the 6-D IC space. One of these diagrams, IC2 versus IC5, is
zoomed to highlight the detailed features in Figure \ref{f7}. The
evolutionary track of an SSP in the IC space, though very complex,
is treatable, because the spectra of a specific Simple Stellar
Population, $SSP^{j}$, is determined by its age, t, and
metallicity, Z, in the frame work of the BC03's model; and in the
same time it can be expressed as a non-negative linear combination
of the 6 ICs,
\begin{equation}\label{eq4}
SSP^{j}(\lambda) = SSP^{j}(t,Z)=\sum_{i=1}^6
c^{j}_{~i}~IC^{i}(\lambda).
\end{equation}
In principle, the correspondence of age and metallicity, (t,Z),
and a 6-dimension array of the 6-IC expanded coefficients,
($c^{j}_{~i},~ i=1,2,...6$), can be found.

The spectrum of any stellar system, e.g., a star cluster or a
galaxy (with no extinction and stellar velocity dispersion), can
also be expressed as a non-negative linear combination of the 6
ICs,
\begin{equation}\label{eq5}
g(\lambda)=\sum_{i=1}^6 g_{i}~IC^{i}(\lambda).
\end{equation}
Thus any stellar system can be decomposed into $N_{*}$ SSPs of
different ages and metallicities,
\begin{equation}\label{eq6}
g(\lambda)=\sum_{j=1}^{N_{*}} s_{j}~SSP^{j}(\lambda)
\end{equation}
where the minimum number of SSP required is $N_{*}\leq 6$. The
coefficients, $s_{j}$ can be determined by solving the following 6
simultaneous linear equations,
\begin{equation}\label{eq7}
g_{i} = \sum_{j=1}^{N_{*}} s_{j} c^{j}_{~i},~~(N_{*}\leq
6,~i=1,2,...,6).
\end{equation}

Note in most cases, the solutions are not unique. We adopt the
following strategy to choose the closest solutions: The $1^{st}$
SSP is chosen from the 74 SSPs described in \S3.2 by minimizing
the Euclidean distance in the 6-D IC space between the SSP and the
galaxy under study. Then the $2^{nd}$ SSP is picked to further
reduce the distance. This procedure is repeated until the
improvement is no longer significant. For a majority of the
galaxies, $N_{*}=2-6$ SSPs are needed to reconstruct the original
spectra. The reliability of this solution is tested in \S5.

\section{Modelling the SDSS Galaxies}

In this section we use the 6 ICs to model the SDSS spectra, which
have comparable spectra resolution with that of GALAXEV. The
spectra of $\sim 2.6\times 10^5$ galaxies in the SDSS DR2
(Abazajian et al. 2004) are fitted to exploit their spectral
features, and in the meantime to test our new technique.

\subsection{Fitting the Galaxy Spectra}

The $\sim 2.6\times 10^5$ galaxies are corrected for the
foreground galactic extinction in the first place using the
extinction curve of Schlegel et al. (1998) and transformed to
their rest frame using the redshift provided by the SDSS
spectroscopic pipeline (Schneider et al. 2002). The derived galaxy
spectra can be fitted by the 6 ICs as:
\begin{equation}\label{eq8}
g^{obs}(\lambda)=A(E_{B-V},\lambda)\sum_{i=1}^6
g_{i}~IC^{i}(\lambda,\sigma_{*}),
\end{equation}
where $g^{obs}(\lambda)$ is the observed spectrum of the galaxy
under study, $IC^{i}(\lambda,\sigma_{*})$ is the $i^{th}$ IC
broadened by convolving with a Gaussian of $\sigma_{*}$ width to
match the stellar velocity dispersion of the galaxy, and
$A(E_{B-V},\lambda)$ is the intrinsic starlight reddening (an
SMC-like extinction curve is assumed, Pei 1992). The final fit is
done through minimizing reduced $\chi^2$, and $E_{B-V}$,
$\sigma_{*}$, and $g_{i}$ (i=1, 2, ..., 6) are non-negative free
parameters. During the fit, the prominent emission lines (Balmer
systems, $H_{n<6}$ and strong forbidden lines such as
$[OII]\lambda3727$, $[OIII]\lambda\lambda4959,5007$,
$[OI]\lambda6300$, $[NII]\lambda\lambda6548,6583$ and
$[SII]\lambda\lambda6716,6731$) are masked out through excluding
the wavelength region (5 \AA~ in width) for each line. Note if
there exists nuclei emission ($\sim 1,200$ objects in the
``galaxy'' catalog of the SDSS DR2), an additional power law and
broader emission line components are added to the fit. Bad pixels
in the spectra flagged by the SDSS pipeline are also excluded. We
then subtract the modelled stellar spectrum and fit emission line
with various Gaussians (Dong et al. 2005). Replacing the masked
emission line ranges with the line fitting results, we reiterate
the above procedures until both the fitting of stellar spectrum
and emission lines are acceptable. For most of the spectra
($\gtrsim 95\%$), two iterations are enough. In Figure \ref{f8}
(lower right panel) we present the reduced $\chi^2/d.o.f$. The
fits are incredibly good for most of the spectra, with only $\sim
2\%$ having $\chi^2/d.o.f$ $>$ 1.5. Representative example of the
fit are displayed in Figure \ref{f9}.

For about ten thousand high quality SDSS spectra, we repeat our
fit using the 20 HS described in \S3. The average contribution of
each HS is plotted in Figure \ref{f10}, which is similar to that
of Figure \ref{f3}. This further confirms that the 6 ICs we
identified are representative enough to embody almost all of the
information in the spectra of any stellar system. Finally we try a
linear least-square fit to all of the $\sim 2.6\times 10^5$ SDSS
galaxies using the 6 ICs with non-negative constraint released.
The contributions of the 6 ICs to each spectrum is shown in Figure
\ref{f11}. It is remarkable that, even without the non-negative
restriction, $\sim 96.2\% $ of the coefficients are non-negative
and $\gtrsim 99.9\% $ of the negative contribution is $\lesssim
2.5\%~$.

\subsection{Results}
The following spectral parameters of galaxies are derived:
\begin{description}
  \item[Starlight Reddening]
Light from Galaxies suffers from various intrinsic reddening due
to different amount of dust. By fitting the optical spectra, we
found that in more than half of the SDSS galaxies, the color
excess of starlight is $\gtrsim 0.1~mag$ (Figure \ref{f8}, upper
left panel). This indicates that reddening must be properly taken
into account while trying to extract the stellar mass and Star
Formation Rate (SFR) from galaxy spectra. In Figure \ref{f12} we
plot starlight extinction obtained using the method presented in
this paper and that estimated according to Balmer decrement,
$H\alpha/H\beta$. It can be seen that starlight extinction is well
correlated with the extinction of nebular emission lines,
suggesting the starlight reddening obtained using EL-ICA method is
reasonable. A best linear fitting reveals
$E^{starlight}_{B-V}=(0.334\pm 0.01)E^{em}_{B-V}$. This slope is
much flatter than that of Calzetti et al. (1994), who obtained a
value of 0.5.

  \item[Stellar Velocity Dispersion]
Similar to Li et al. (2004), stellar velocity dispersion,
$\sigma_{*}$, can be given directly by the fit (Figure \ref{f8},
upper right panel). The results from two methods are consistent.

  \item[Stellar Population and Metallicity]
Another advantage of the EL-ICA method over PCA-based approach is
that light/mass-weighted ages, metallicities, and stellar mass can
be derived from integrated galaxy spectra, since the galaxy
spectrum is non-negative combination of the spectra of 6 ICs or
$N_{*}\leq 6$ SSPs. The results are also displayed in Figure
\ref{f8}. Our measurement of stellar mass is compared with that of
Kauffmann et al. (2003) in Figure \ref{f13}. In fact, stellar
population can be revealed with higher resolution than mere
average age. Figure 14 shows the modelled star formation histories
of the same 3 SDSS galaxies displayed in Figure 9, expressed as
light and mass fractions versus age. Furthermore, as the recovered
number of ICs, in turn the number of the SSPs needed to model the
spectrum of a galaxy, is no more than six, the probability of
over-fit can be greatly reduced.

\end{description}

It is interesting to inspect the distribution of the observed
galaxy spectra in the IC space. We randomly select $\sim 1,000$
galaxies with median $S/N\gtrsim 30$ from the SDSS DR2 and plot
them in Figure \ref{f6} and \ref{f7}. Most of the galaxies are
enveloped by the SSPs. This is consistent with the assumption that
the star formation histories of any galaxy can be expressed as a
sum of discrete bursts. Absorption-line galaxies tend to
distribute in the surface of the envelop formed by the SSPs, and
they are well separated from emission-line galaxies. This is
because absorption-line galaxies are passively evolving ones,
which only contain old and/or middle-aged SSP(s), while
emission-line (star-forming) galaxies in the local universe often
comprise both young and old stellar populations. We note that
galaxies with larger contribution of ``early-type'' ICs,
corresponding to young populations, tend to have stronger emission
lines. This indicates that, 1) spectral classification of galaxies
can be made using the EL-ICA method; and 2) the optical stellar
spectrum of a galaxy can serve as a star formation rate indicator.
Using the $N_{*}\leq 6$ recovered SSP spectra, the near-IR SED can
also be estimated by extrapolating from the optical data. We leave
this issue for future studies.

\section{Reliability of the Derived Parameters}

We perform extensive tests to estimate the reliability of the
parameters yielded by our galaxy spectra modelling. For this
purpose, 59400 synthetic galaxy spectra are created from the
library of the BC03 SSP following the steps below:
\begin{itemize}
  \item
We first pick up 60 SSP spectra, with 10 different ages (0.005, 0.01, 0.02,
0.06, 0.2, 0.6, 1.4, 5, 10, 15 Gyr) for each of the 6 metallicities.
  \item
6 out of the 60 SSP spectra are randomly selected, one for each
metallicity. The 6 spectra are then normalized at 5500 \AA\ and
combined with random weights to create one synthetic spectrum. The
total weight of the 6 SSPs is normalized to 1. Repeating this step
we get a total of 60 synthetic spectra.
  \item
The 60 spectra are then broadened by convolving with Gaussian
function with stellar velocity dispersion of
$\sigma_{*}=30m~km~s^{-1}$, m=1, 3, ..., 15.
  \item
Different reddening are applied to enlarge the sample to 9,900
spectra, with $E_{B-V}=0.1n~mag$ , n=0, 1, ..., 11.
  \item
Gaussian noise is finally added to yield spectra sets with S/N=10, 15,
20, 30, 40, 50 respectively.
\end{itemize}
In this way 59,400 synthetic galaxy spectra are created, with wide
coverage of ages and metallicities, 15 stellar velocity
dispersions, 11 color excess, and 6 S/N levels. The synthetic
galaxy spectra in the wavelength range of $3322-9200$ \AA\ are
fitted using Equation \ref{eq8} as we model the galaxies in the
SDSS DR2 and the input parameters are recovered simultaneously.
During the fit, we mask the same potential emission line regions
used to fit the real galaxies, although it is not required because
the synthetic spectra do not contain emission lines. This makes
our test results directly comparable to the results from the real
spectra:

\begin{description}
  \item[Starlight Reddening]
There is no systematic discrepancy between the recovered and the
input starlight reddening. In Figure \ref{f15}, we plot the rms of
the measured color excess. The uncertainty is not sensitive to the
S/N ratio of synthetic galaxy spectra, provided $S/N\gtrsim 15$
per pixel. This is understandable because S/N ratio does not
significantly affect the SED.
  \item[Stellar Velocity Dispersion]
The recovered and the input stellar velocity dispersion,
$\sigma_{*}$, are plotted in Figure \ref{f16}. The disagreement
between the two is $\lesssim 30~km~s^{-1}$ for $S/N>10$ and the
measurement uncertainty decreases rapidly with increasing S/N
ratio. For typical spectral quality of the SDSS galaxies ($S/N\sim
20$), $1~\sigma $ error is $\lesssim 20~km~s^{-1}$.

  \item[Stellar Populations, Metallicities and stellar mass]
Using the EL-ICA method, one can decompose a galaxy spectrum into
$N_{*}\leq 6$ SSPs of different ages and metallicities. In turn,
stellar mass, light (or mass)-weighted average age and metallicity
can be derived. The comparisons between the recovered and input
values are displayed in Figure 17-19. The results indicate that
the age-metallicity degeneracy can at least be partly broken and
star formation histories be estimated. Figure \ref{f20} shows that
the star formation history can be resolved with higher resolution
than average age. It can be seen that, though the star formation
histories of a galaxy cannot be exactly recovered due to the
age-metallicity degeneracy, they can be estimated with meaningful
accuracy.

\end{description}

\section{Comparison with PCA}

PCA is a classic technique in statistical data analysis, feature
extraction, and data compression. It has been successfully used in
studies of the multivariate distribution of astronomical data
(e.g., Efstathiou \& Fall 1984; Connolly et al. 1995; Folkes et al. 1996;
Glazebrook et al. 1998; Bromley et al. 1998;
Folkes et al. 1999; Madgwick et al. 2002; Li et al. 2005). Given a
set of multivariate observations, the basic PCA problem is to find
a small set of variables with less redundancy that can give as
good representation as possible. Though the goal of PCA is similar
to that of ICA, the method used to get rid of redundancy is
different: In PCA the redundancy is measured by correlations
between observed data, while the much richer concept of
independence is used in ICA. The advantage of ICA has been
mentioned in \S2.1. However, using only the correlations in PCA
has likewise its advantage that the analysis can be based on
2nd-order statistics only. It would be interesting to make
comparison between the two techniques. In this section, we apply
PCA to analyze the same data set as we have done in the previous
sections using ICA.

We first apply PCA to the truncated SSP spectra in GALAXEV
mentioned in \S3.1, which are also normalized at 5500 \AA. All of
the 1326 spectra in the library, instead of a subsample, are
analyzed this time since PCA converges much faster than EL-ICA. We
subtract the mean from the spectra before PCA as Madgwick et al.
(2002) and adopted the mathematical formulation of PCA given in
Folkes et al. (1999). The PCA analysis generates a set of
eigenspectra, which are denote as $PC^{i}(\lambda)$ (i=1,2,3,...,
ordered by their relative importance measured as variance), with
the main spectral features of the whole library concentrated in
the first few. Indeed, we find that the cumulative contribution of
the first three PCs to the total variance amounts to 99.7\%, and
the significance of each successive PC drops off sharply. We
determined the number of PCs required to represent GALAXEV as
follows. Initially, the SSP spectra in GALAXEV were fitted by the
first three PCs as Equation \ref{eq4} substituting PCs to ICs and
releasing the non-negative restriction on $c^{j}_{~i}$. By adding
successively the next PC to the model, we calculate the
significance of the improvement to the fit using the F-test,
\begin{eqnarray}\label{eq9}
\alpha_F & = & \int_F^\infty dF p (F|\Delta P, N-P_1) =
I_{\frac{N-P_1}{N-P_1+\Delta P \cdot F}}
\big(\frac{N-P_1}{2},\frac{\Delta P}{2}\big),
\end{eqnarray}
where, the $F$-statistics $ F=\frac{\Delta \chi^2 / \Delta P}
{\chi_1^2 / (N-P_1)}$, $P_1$ and $\Delta P=1$ are the numbers of
thawed parameters of the previous model and of the additional
freely varying parameters in the current model, $I$ the incomplete
beta function. Adopting a critical significance of
$\alpha_F=0.05$, we find that more than 80\% SSP spectra can be
well-fitted using the first 9 PCs (see Figure. \ref{f21}), which
are then chosen as our final spectral templates (also denoted as
PCs).

Then we fit all the galaxies in the SDSS DR2 using Equation
\ref{eq8} in the same way as described in \S4.1, substituting the
nine PCs to the six ICs and releasing the non-negative restriction
on $g_{i}$. It can be seen from Figure \ref{f22} (upper panel)
that the distribution of reduced $\chi^2$ is similar to that of
EL-ICA models, indicating that starlight in observed galaxy
spectra can also be well reproduced using the nine PCs. Like
EL-ICA method, we can also obtain stellar velocity dispersion as a
by-product. We compare the measured values of $\sigma_{*}$ using
the two methods and find that they agree with one another with in
the measurement uncertainty (Figure \ref{f22}, lower panel). We
plot in Figure \ref{f23} the starlight extinction as determined by
PCA method against the extinction of nebular emission lines for
the same $\sim 500$ HII galaxies as in \S4.2. The two values are
also correlated but the scatter is a little larger than Figure
\ref{f12}. This result is consistent with that of Li et al.
(2005). These authors also tried to determine starlight reddening
using PCA based approach and found the typical uncertainty is
$\sim 0.2~mag$, much larger than that yielded by the EL-ICA method
($\sim 0.06~mag$, see \S5.1). Due to the fact that we cannot force
PC spectra non-negative (they must be perpendicular with one
another) and the combination coefficients, we are unable to obtain
the stellar contents of galaxies.

\section{Summary}

We employ EL-ICA to analyze stellar system spectra and compress
the BC03 stellar spectral library into 6 non-negative ICs. In
consequence, the spectrum of a stellar system can be compressed to
a vector in the 6-dimension space spanned by the 6 ICs, which
redescribes the original data in a very condense form. These 6 ICs
are used as templates to model the spectra of normal galaxies in
the SDSS DR2. Important spectral parameters, such as starlight
reddening, stellar velocity dispersion, stellar contents and star
formation histories, can be obtained simultaneously. Extensive
tests show that satisfactory accuracy can be achieved to these
parameters for the typical spectral quality of the SDSS. The ICA
model does not depend on previous knowledge. Using the EL
algorithm to ICA model, we can also force the ICs and the
combination coefficients non-negative and find the simplest
interpretation of the observed galaxy spectra. The physical
meaning of the ICs and the separation of the observed galaxy
spectra can be easily understood. All the features of stellar
spectrum are taken into account in the fit, hence the conundrum of
over-fit is greatly mitigated. The method presented in this paper
open a new way to model galaxy spectra. Its robustness and high
efficiency make it applicable to extract the spectral
characteristics of galaxies observed by modern large sky area
surveys.

%%%%%%%%%%%%%%%%%%%%%%%%%%%
\acknowledgements

We thank the anonymous referee for useful suggestions that lead to
improvement of the paper. This work was supported by Chinese NSF
through NSF10233030, the Bairen Project of CAS, and a key program of
Chinese Science and Technology Ministry. This paper has made use of
the data from the SDSS. Funding for the creation and the distribution
of the SDSS Archive has been provided by the Alfred P. Sloan Foundation,
the Participating Institutions, the National Aeronautics and Space
Administration, the National Science Foundation, the U.S.
Department of Energy, the Japanese Monbukagakusho, and the Max
Planck Society. The SDSS is managed by the Astrophysical Research
Consortium (ARC) for the Participating Institutions. The
Participating Institutions are The University of Chicago,
Fermilab, the Institute for Advanced Study, the Japan
Participation Group, The Johns Hopkins University, Los Alamos
National Laboratory, the Max-Planck-Institute for Astronomy
(MPIA), the Max-Planck-Institute for Astrophysics (MPA), New
Mexico State University, Princeton University, the United States
Naval Observatory, and the University of Washington.

\clearpage

\begin{figure}
\epsscale{0.8} \plotone{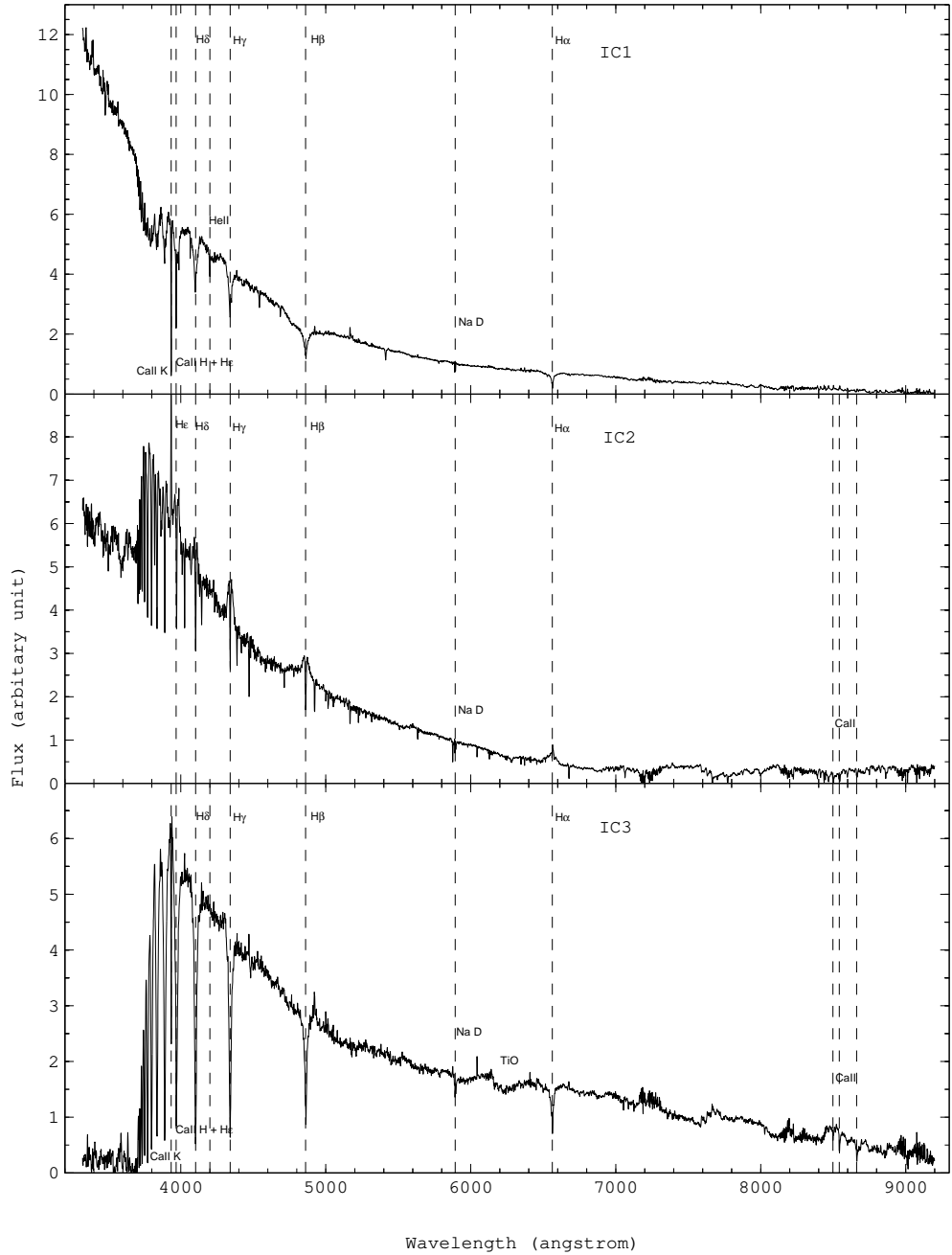} \caption{The spectra of the 6
ICs ordered by their ``spectral types'' (from ``early'' to
``late'', see detailed description in the text). The prominent
spectral features are labelled.} \label{f4}
\end{figure}

\begin{figure}[tbp]
\figurenum{4} \epsscale{0.8} \plotone{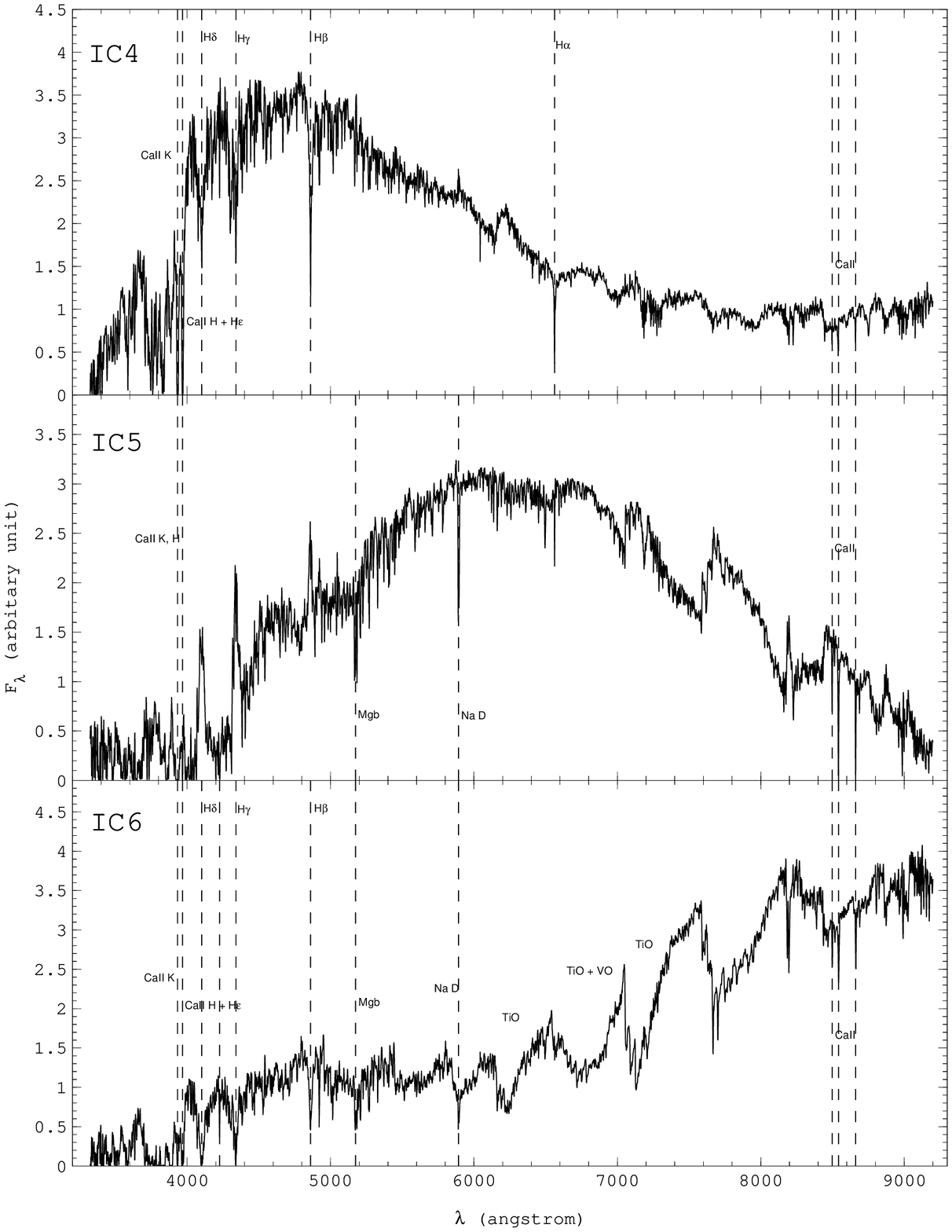}
\caption{Continued...}
\end{figure}

\begin{figure}
\epsscale{0.8} \figurenum{5} \plotone{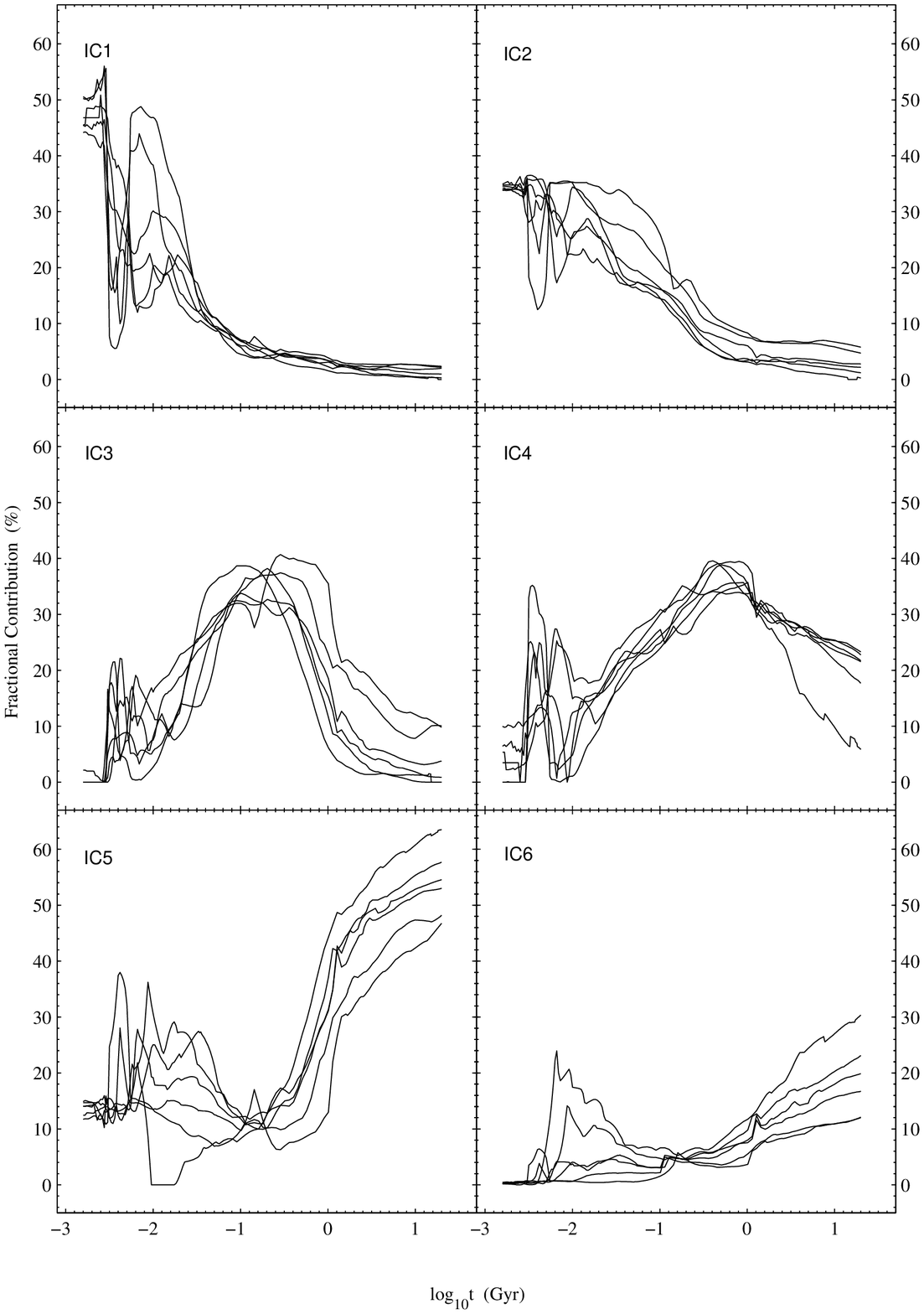} \caption{The
projection of the SSP spectra on the $ICi^{th}$ versus age. Each
line denotes one metallicity. } \label{f5}
\end{figure}

\begin{figure}
\epsscale{0.9} \figurenum{6} \plotone{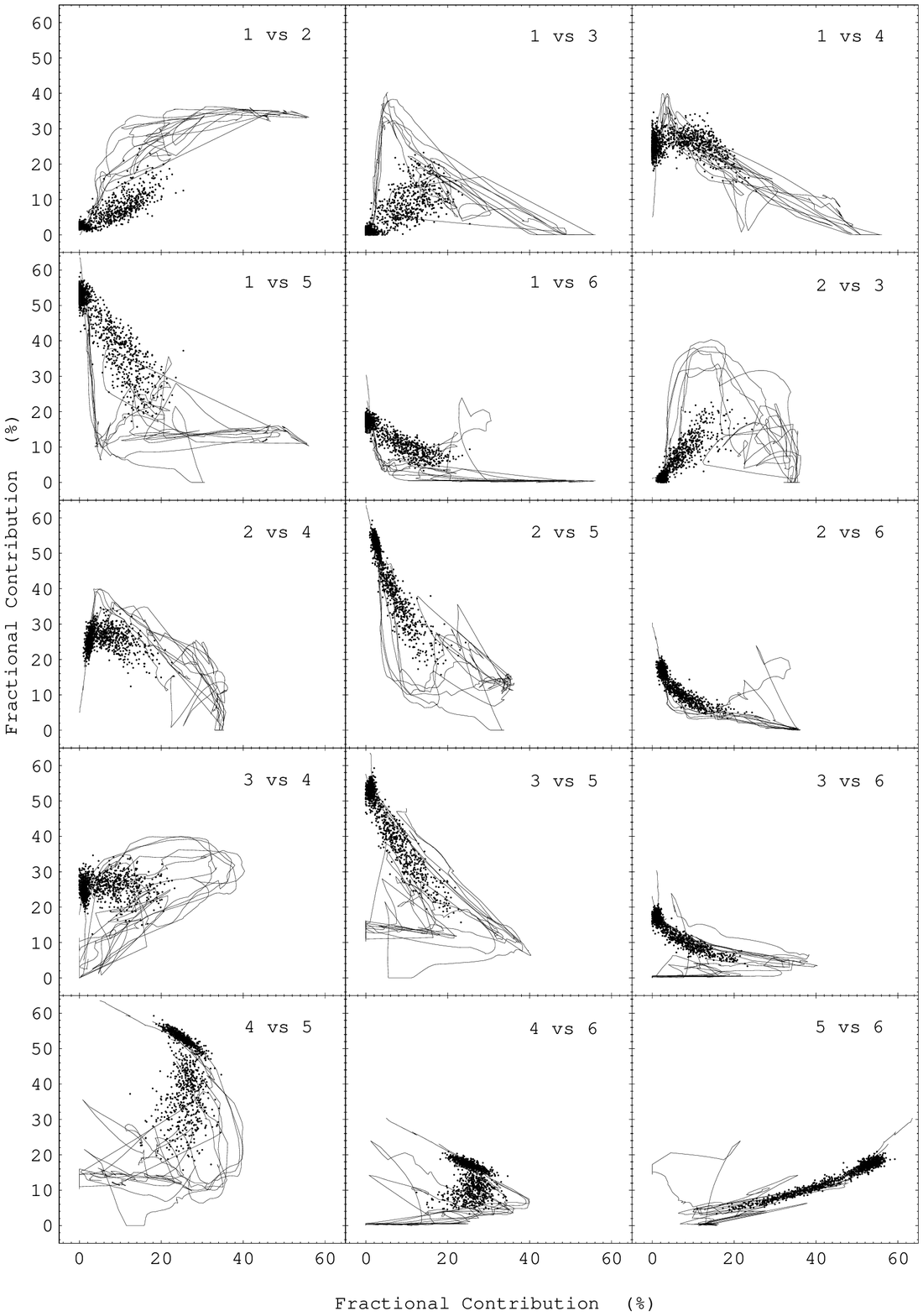} \caption{The
projection of the SSP (lines) and the SDSS galaxy spectra (dots)
on the IC (x-axis) vs IC (y-axis) plane. The $\sim 1,000$ high
spectral quality galaxies are randomly selected from the SDSS DR2.
} \label{f6}
\end{figure}

\begin{figure}
\epsscale{0.8} \figurenum{7} \plotone{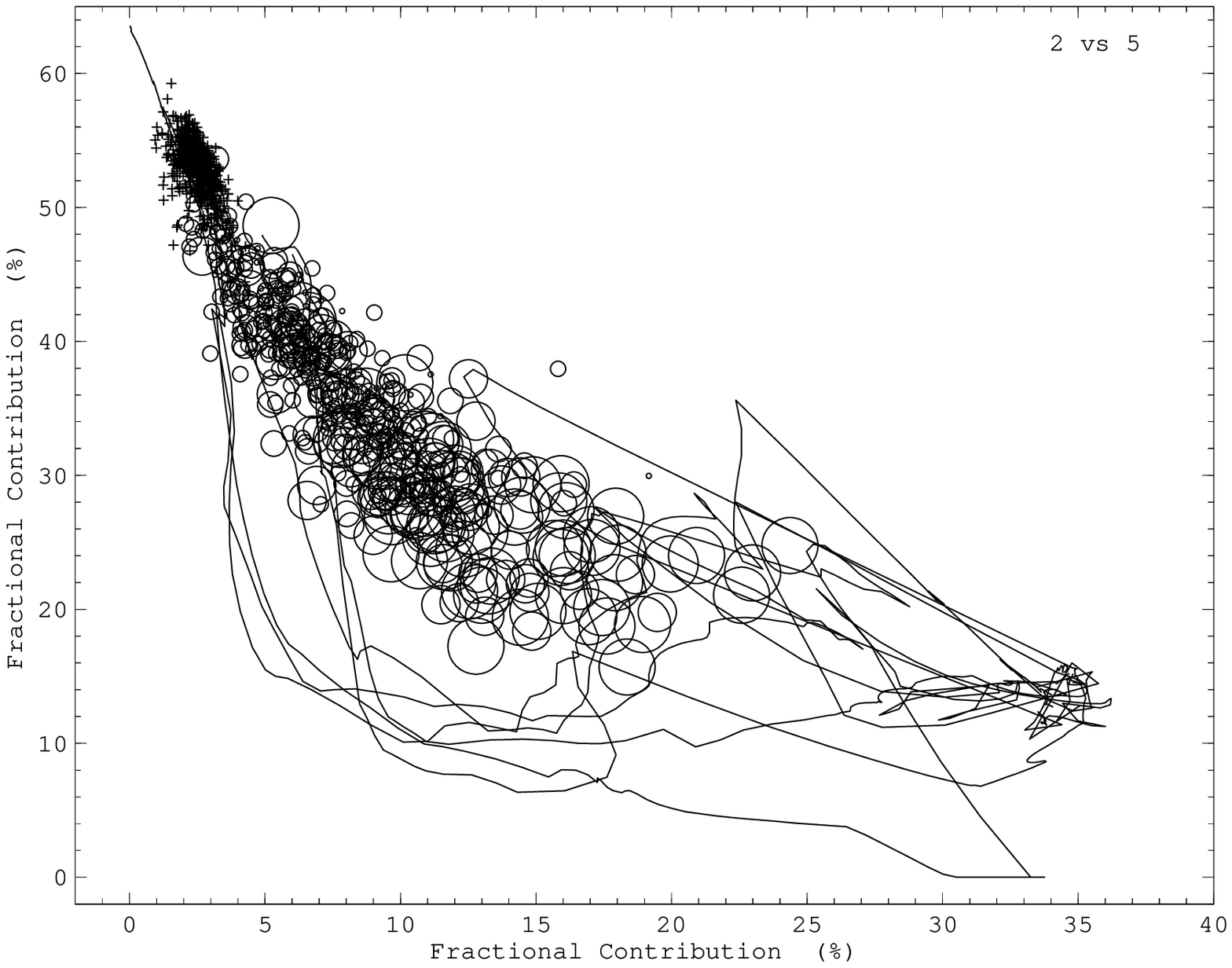} \caption{A close
look at the center penal of Figure \ref{f6} (IC2 (x-axis) vs IC5
(y-axis)), but with the SDSS absorption line galaxies denoted as
cross and emission line galaxies denoted as circle of size
increasing with the equivalent width of $H\alpha$, $EW(H\alpha)$.
It is obvious that the emission line galaxies are well separated
from absorption line galaxies in this diagram. Also note that
galaxies with larger contribution of IC2, corresponding to young
populations, tend to have stronger emission lines. } \label{f7}
\end{figure}

\begin{figure}
\epsscale{0.8} \figurenum{8} \plotone{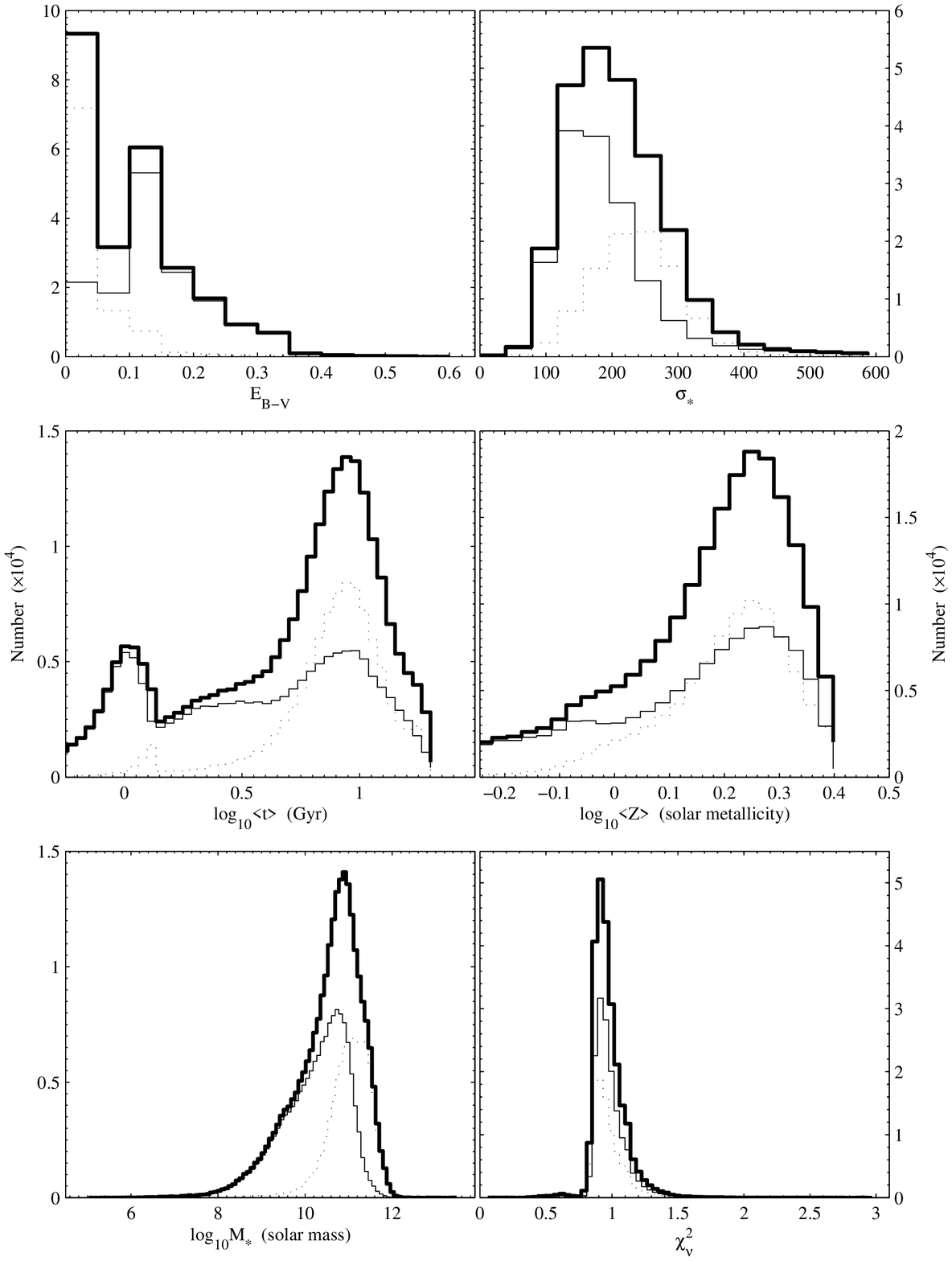} \caption{The
distribution of starlight reddening ($E_{B-V}$, upper left panel),
stellar velocity dispersion ($\sigma_{*}$, upper right panel),
light-weighted average age ($<t>$, middle left panel),
light-weighted average metallicity ($<Z>$, middle right panel),
stellar mass ($M_{*}$, lower left panel), and reduced $\chi^2$
($\chi^2_{\nu}=\chi^2/d.o.f$, lower right panel) for model fits of
all the $\sim 2.6\times 10^{5}$ galaxies in the SDSS DR2. Total
objects are denoted as thick solid lines, absorption line galaxies
as dot lines, and emission-line galaxies, including HII galaxies,
AGN and composite objects as thin solid lines. Aperture effect is
not corrected. } \label{f8}
\end{figure}

\begin{figure}
\epsscale{0.8} \figurenum{9} \plotone{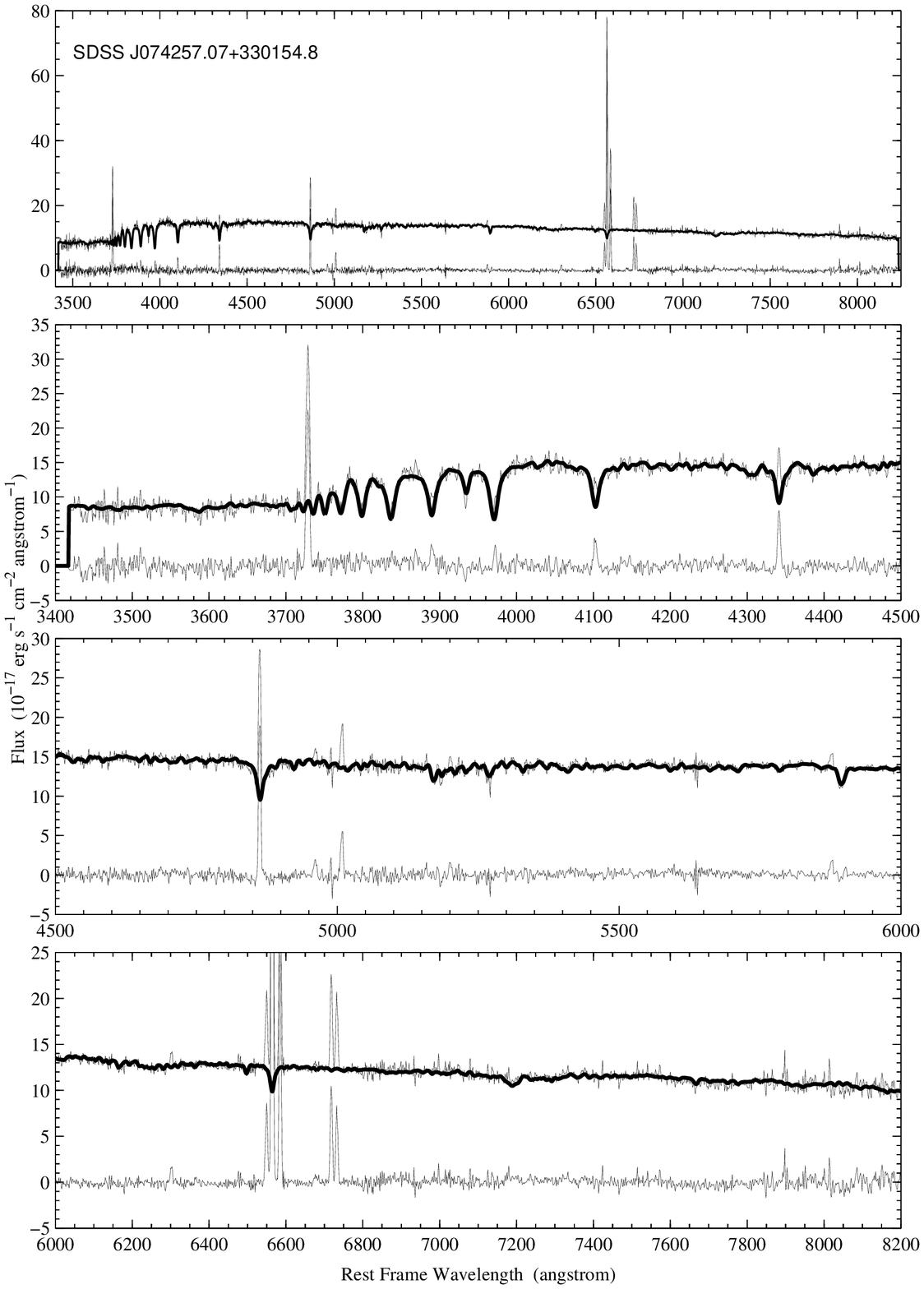}
\caption{Examples of nonnegative linear least-square fits of the
galaxy spectra in the SDSS DR2. The fits are yielded using the 6
ICs derived through the EL-ICA method. Emission lines, when
present, are filtered to perform the fit. The upper panel displays
the model fit as a whole and detailed features are highlighted in
the lower three panels. In each panel, the upper two lines denote
the observed spectrum (thin curves) and the modelled spectrum
(thick curves). The lower thin line denotes the residual (or pure
emission line) spectrum. The apparent mismatch around 6900 \AA~ is
due to the residual telluric absorption in the spectra of the BC03
SSPs, which we use to derive the templates. } \label{f9}
\end{figure}

\begin{figure}
\epsscale{0.9} \figurenum{9} \plotone{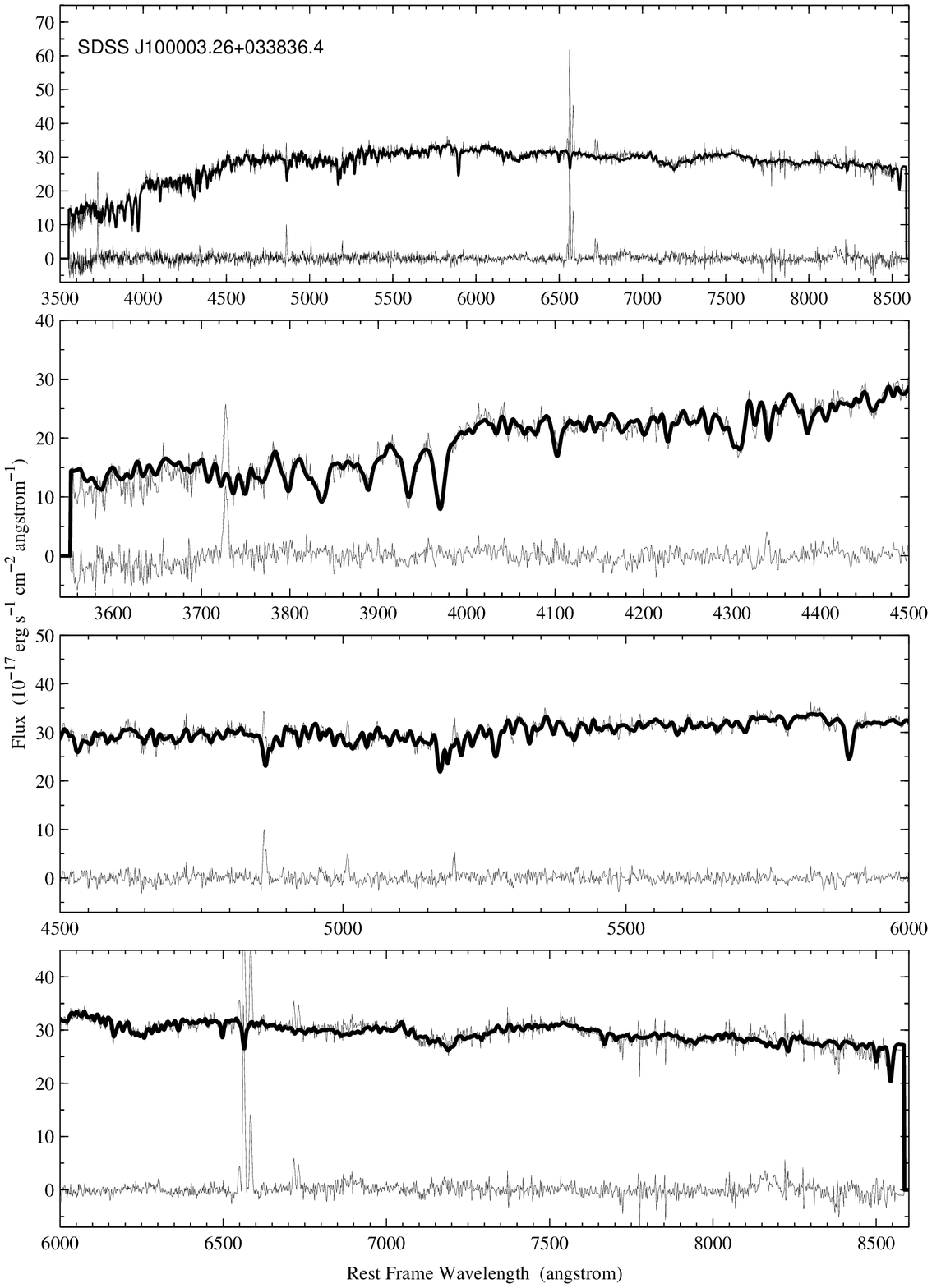}
\caption{Continue... } \label{IC_a}
\end{figure}

\begin{figure}
\epsscale{0.9} \figurenum{9} \plotone{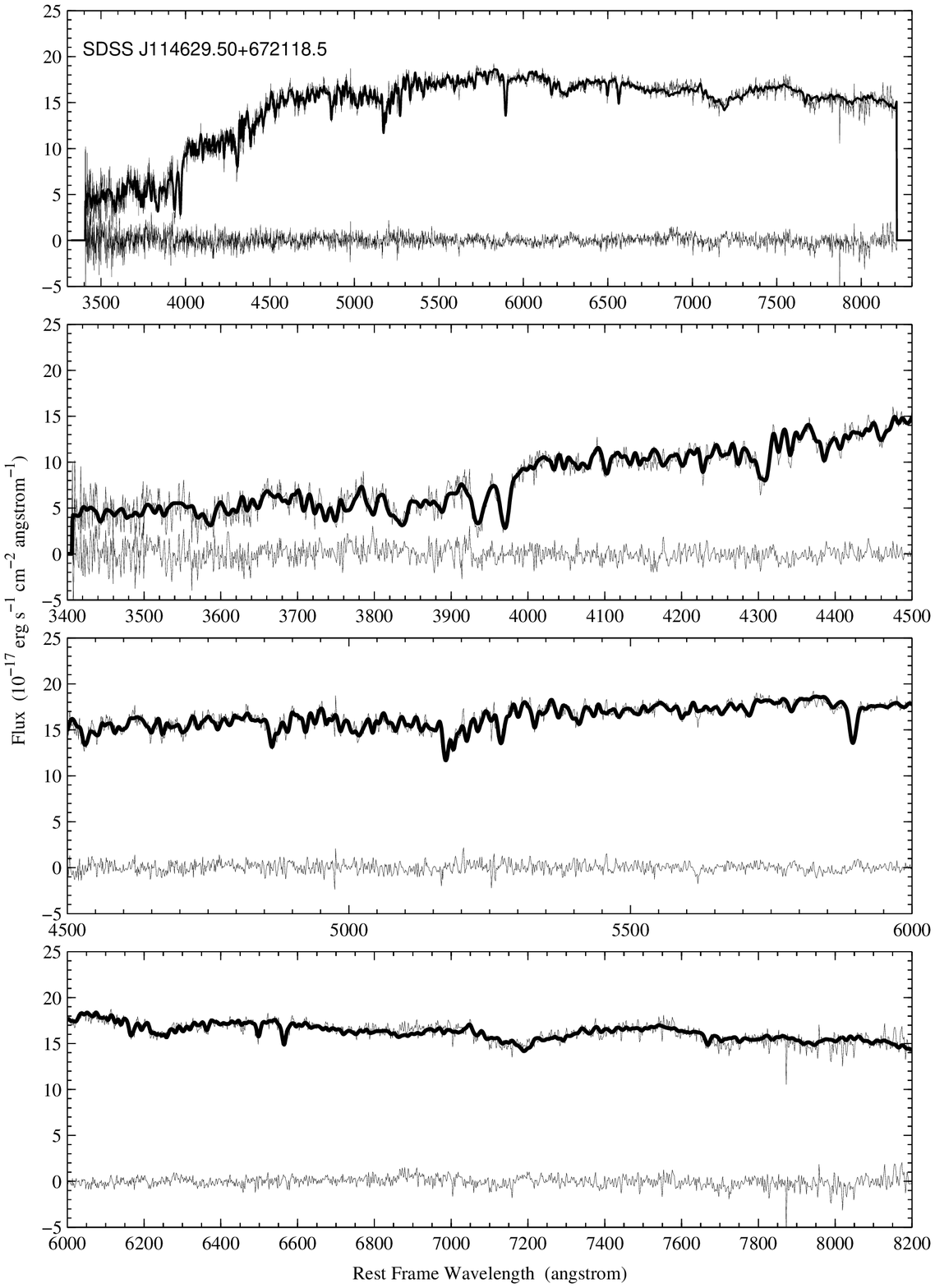}
\caption{Continue... } \label{IC_b}
\end{figure}

\begin{figure}
\epsscale{0.8} \figurenum{10} \plotone{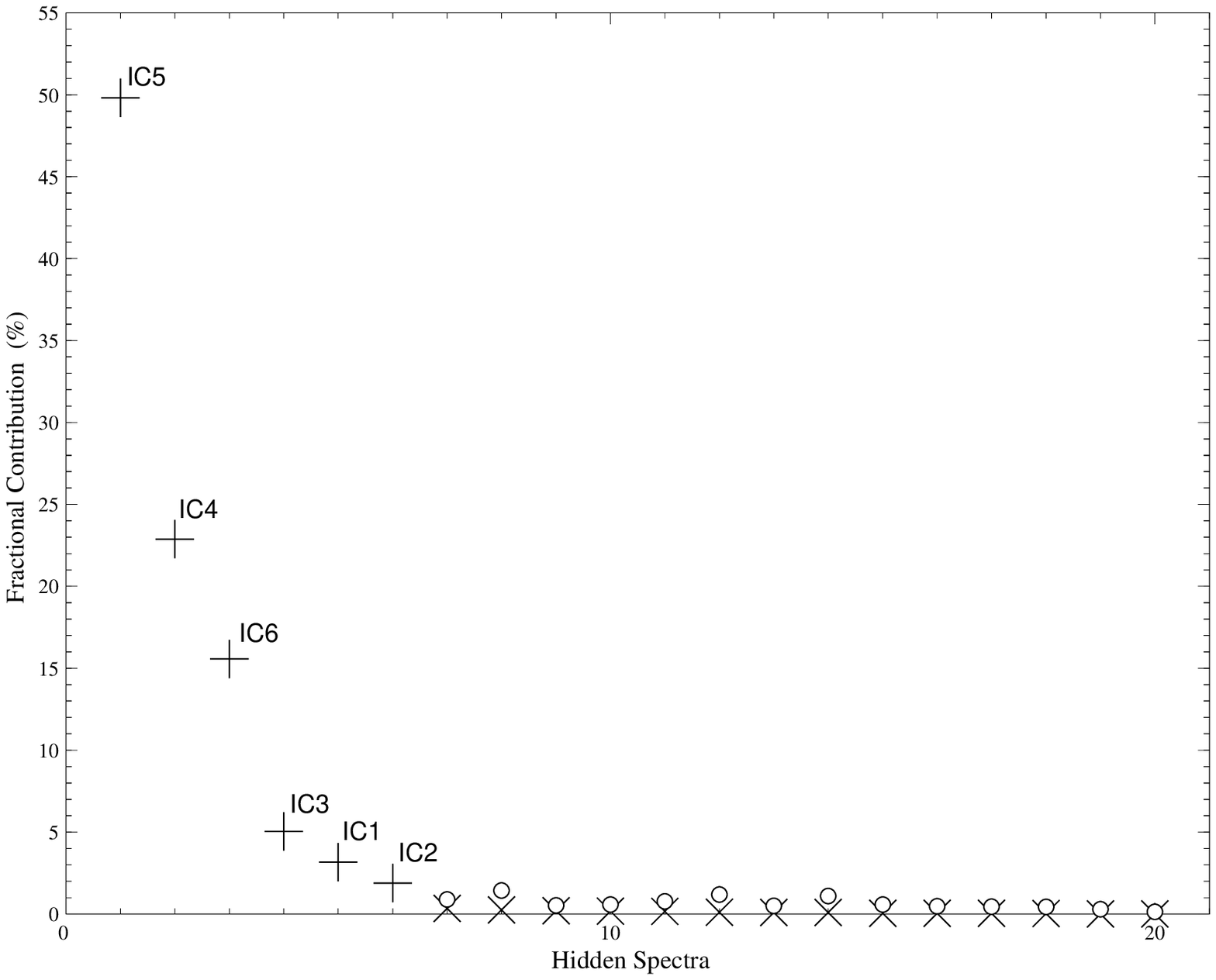} \caption{Same
as Figure \ref{f3} but for $\sim 10^4$ high quality spectra of the
SDSS galaxies. The total contribution of the 6 ICs (the first 6
Hidden Spectra) is $98.36\%$, which is even higher than their
total contribution to the BC03 SSP spectra. The ICs are ordered by
their spectral types, as in Figure \ref{f3} and \ref{f4}.  }
\label{f10}
\end{figure}

\begin{figure}
\epsscale{0.9} \figurenum{11} \plotone{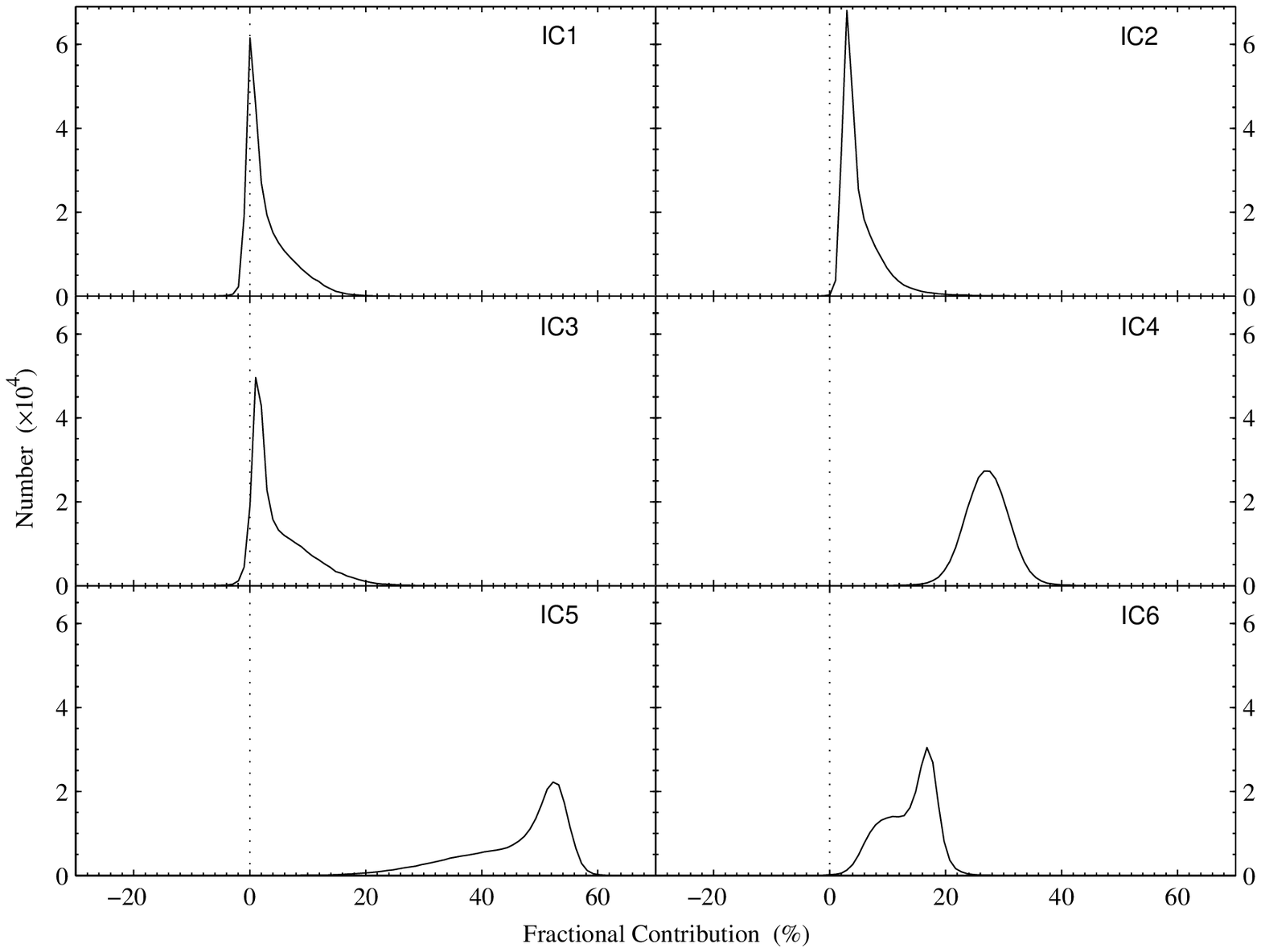}
\caption{Distribution of fractional contribution of the 6 ICs to
the $\sim 2.6\times 10^5$ galaxies in the SDSS DR2, if releasing
the non-negative requirement. It is remarkable that only a
majority of the contribution is non-negative and the amplitude of
negative contribution is negligible.} \label{f11}
\end{figure}

\begin{figure}
\epsscale{0.8} \figurenum{12} \plotone{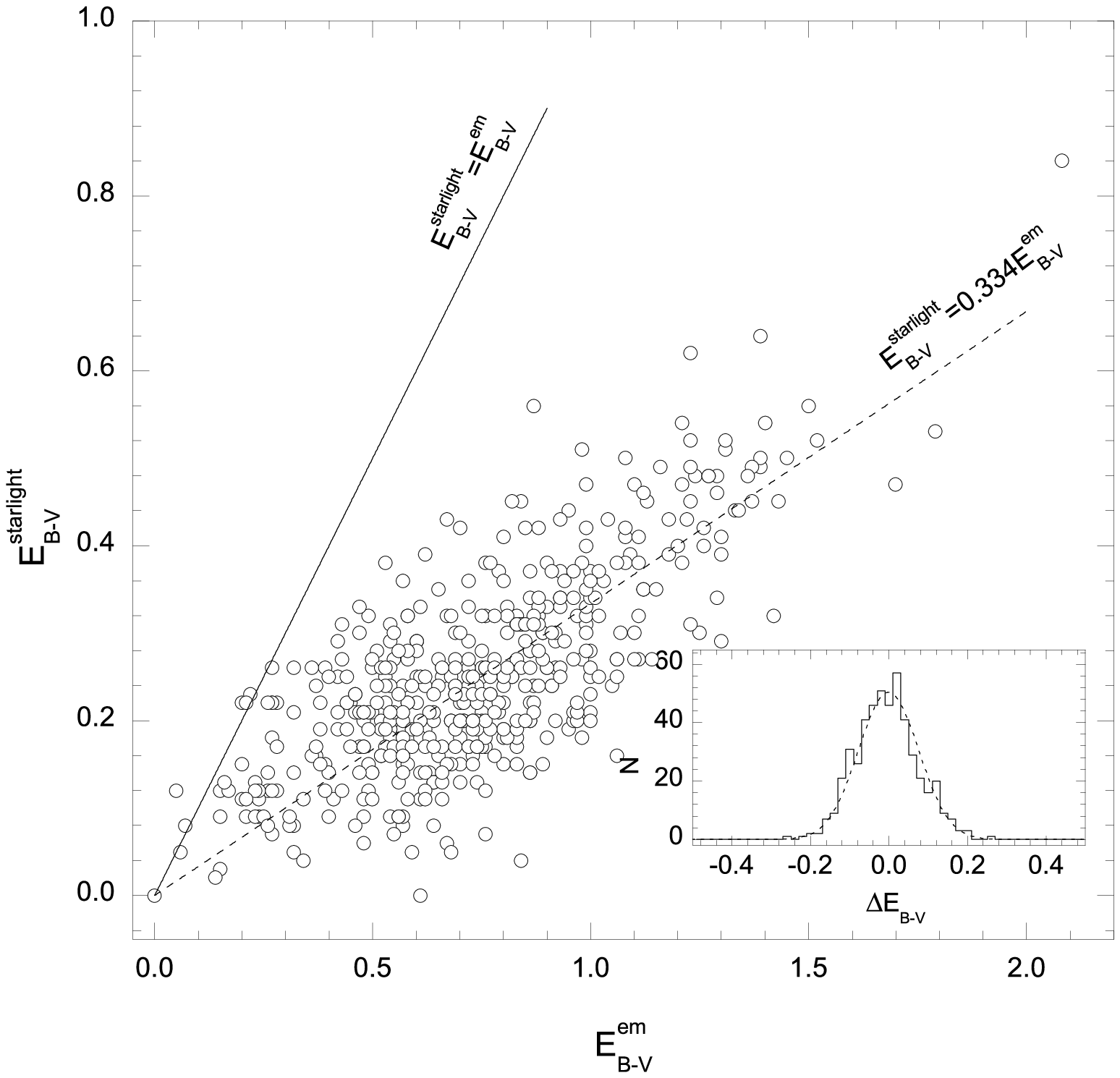} \caption{Mean
color excess of starlight, $E_{B-V}^{starlight}$, obtained by our
continuum model fitting versus that of emission lines,
$E_{B-V}^{em}$, estimated by Balmer decrement, $H\alpha/H\beta$,
for $\sim 500$ HII galaxies with high quality spectra (median S/N
ratio $\gtrsim 30$ and $H\beta$ flux $>10~ \sigma$) arbitrarily
selected from the SDSS DR2. $E_{B-V}^{em}$ is estimated assuming
an intrinsic Balmer decrement, $H\alpha/H\beta =2.87$ and the SMC
extinction curve. The dash line is best linear fitting. The inset
plot shows the deviation of the measured values and the best fit
line, $\Delta E_{B-V}= E_{B-V}^{predicted}-E_{B-V}^{measured}$.
The dashed line is Gaussian with $\sigma=0.076~mag$. } \label{f12}
\end{figure}

\begin{figure}
\epsscale{1.0} \figurenum{13} \plotone{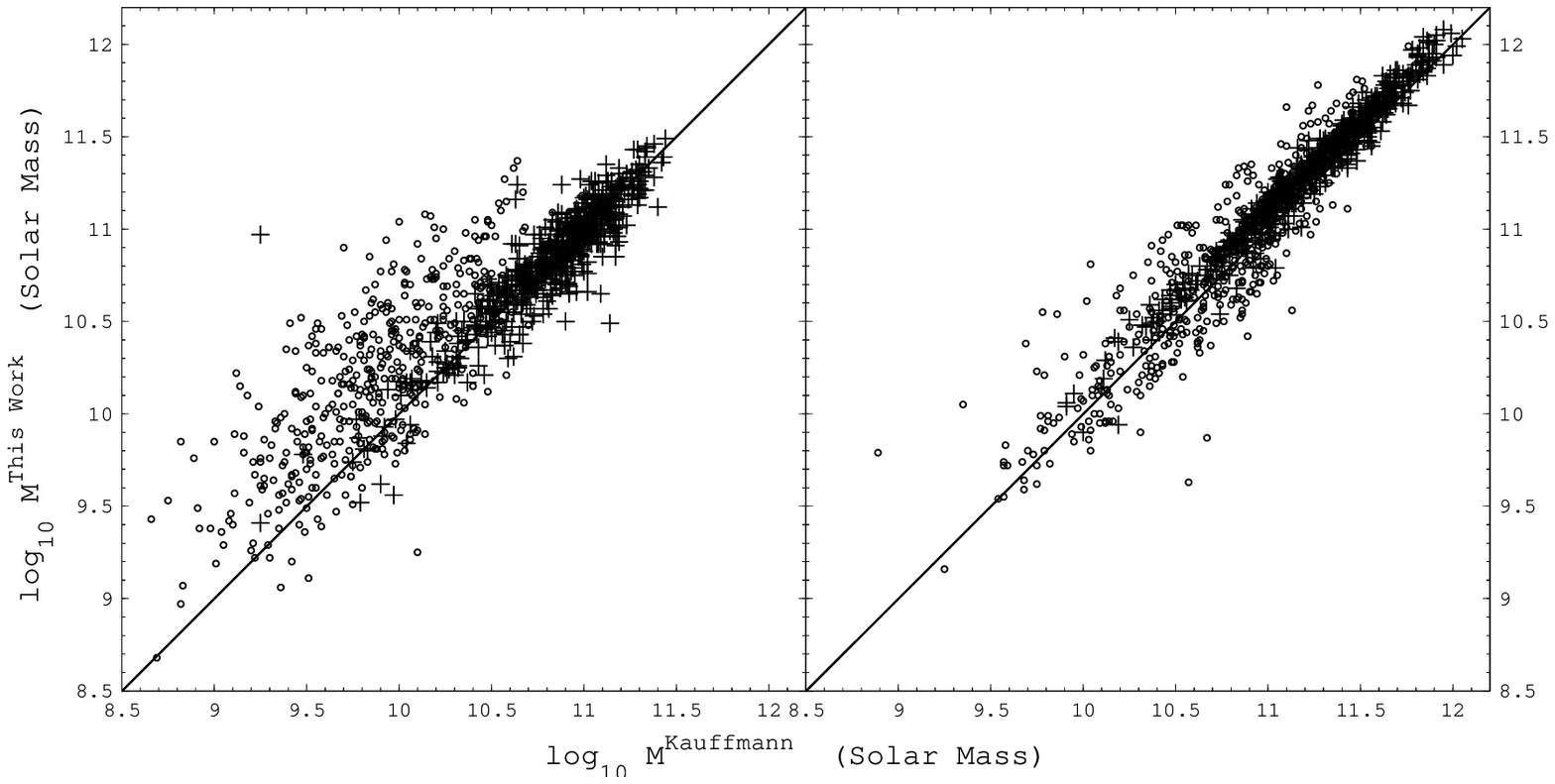} \caption{Our
measurement of stellar mass compared with that of Kauffmann et al.
(2003). Only $\sim 1,000$ typical SDSS galaxies are shown for
clarity. Emission line galaxies are denoted as circle and
absorption line galaxies as cross. In the left panel, stellar mass
of our measurement is corrected for intrinsic extinction but not
for aperture effect, and value of Kauffmann et al. was not
corrected for both of dust reddening and aperture effect. Stellar
mass is corrected for both of aperture effect and dust reddening
in the right panel. Our measurement agrees well with Kauffmann et
al. The agreement is better for absorption line galaxies than for
emission line galaxies, in which dust reddening is often more
serious. } \label{f13}
\end{figure}

\begin{figure}
\epsscale{0.8} \figurenum{14} \plotone{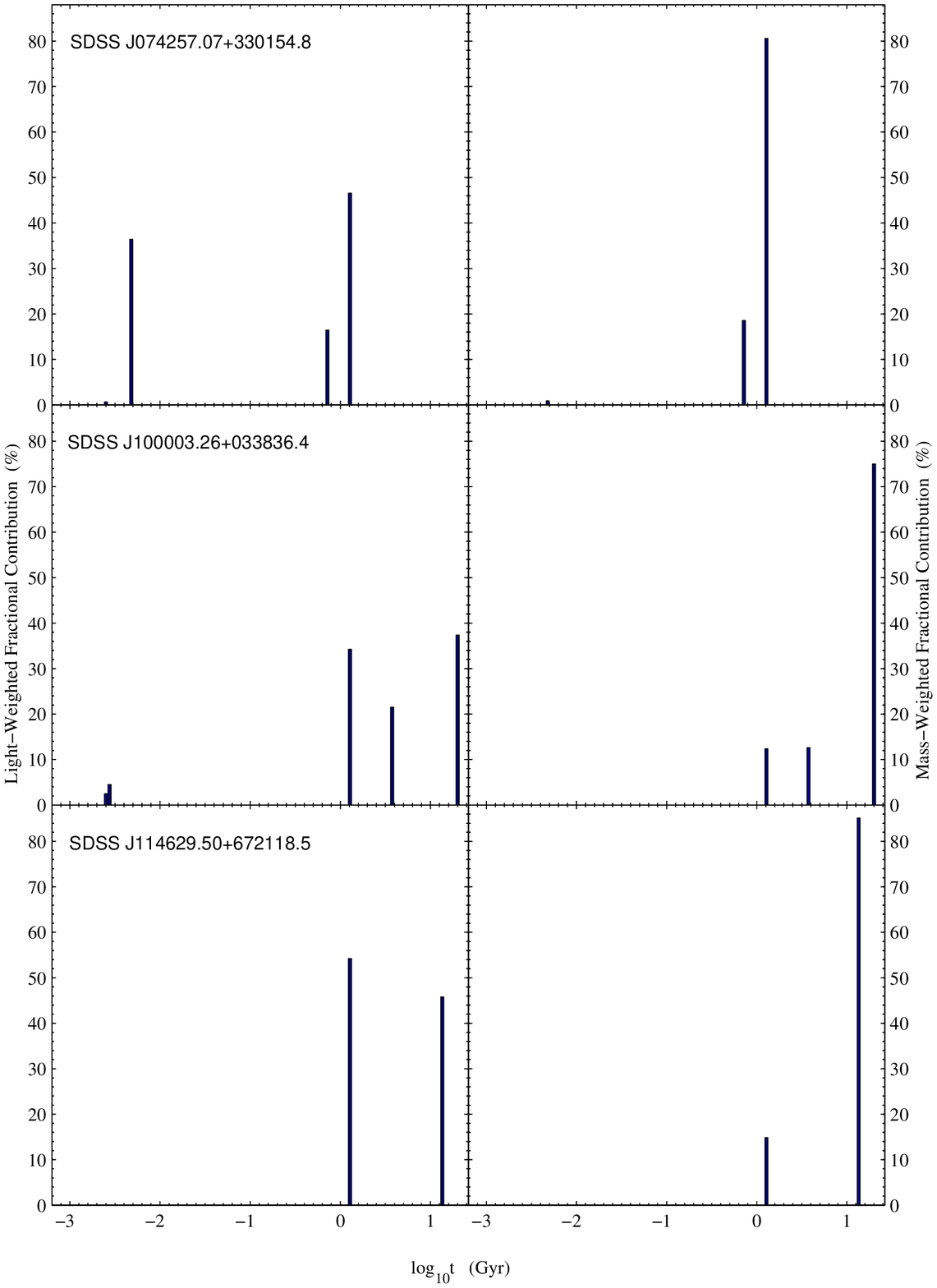} \caption{The
modelled star formation histories of the 3 SDSS galaxies as in
Figure \ref{f9}, expressed as light (left panels) and mass (right
panels) fractions versus ages. } \label{f14}
\end{figure}

\clearpage

\begin{figure}
\epsscale{1.0} \figurenum{15} \plotone{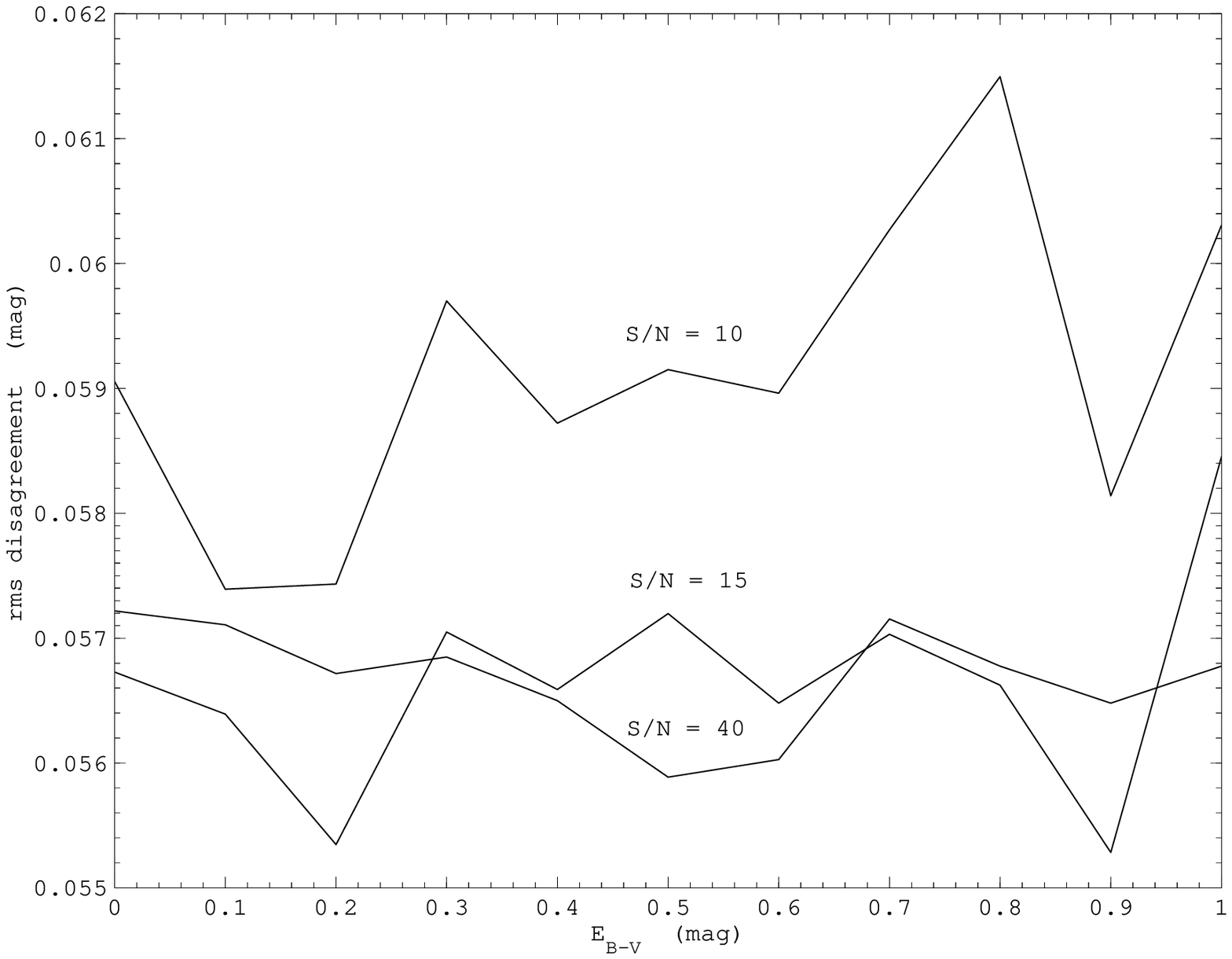} \caption{The
uncertainty (rms) of the measured starlight extinction for the
artificial spectra with different signal-to-noise ratios. }
\label{f15}
\end{figure}

\begin{figure}
\epsscale{1.0} \figurenum{16} \plotone{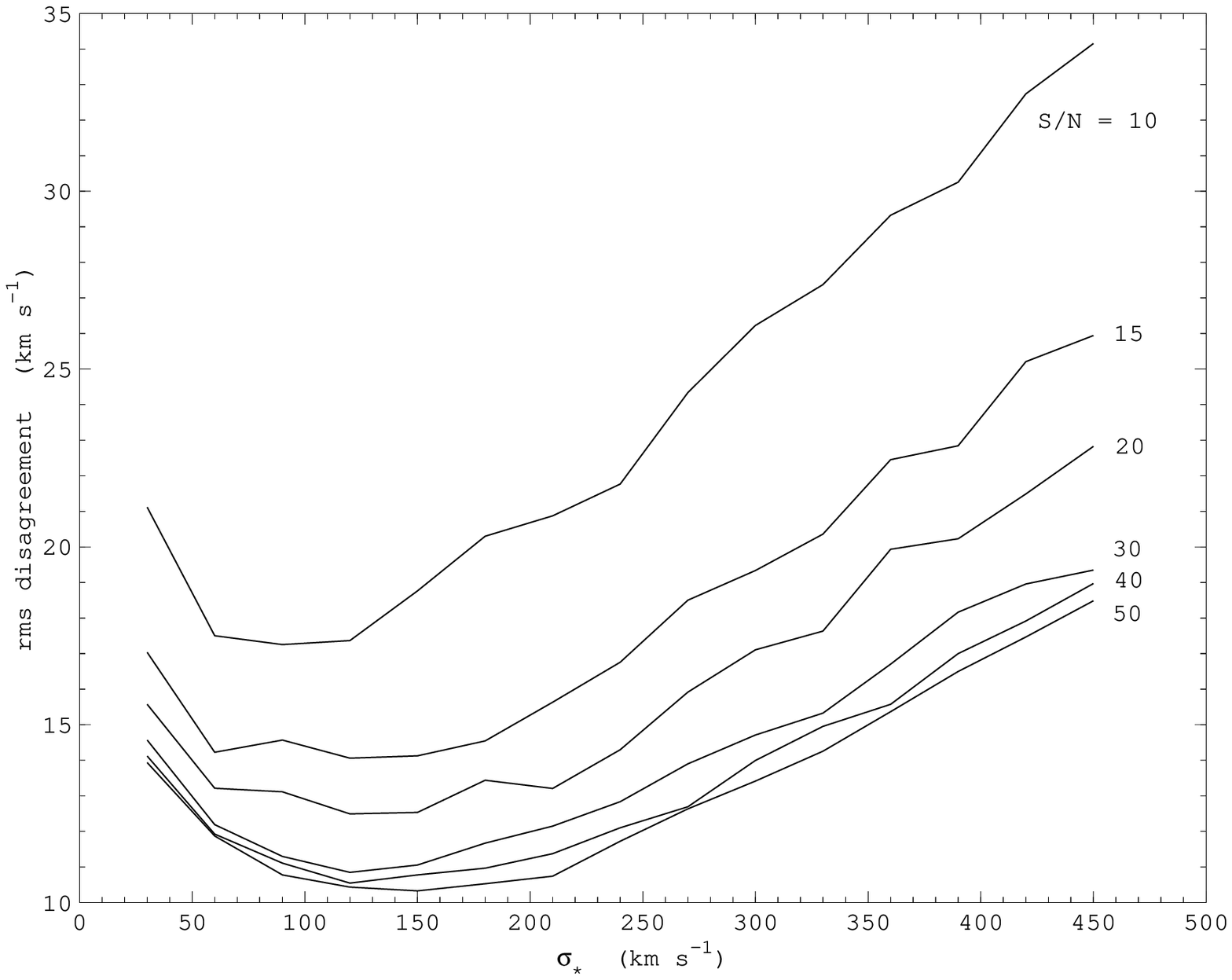} \caption{The
uncertainty (rms) of the measured stellar velocity dispersion for
the artificial spectra with different signal-to-noise ratios. }
\label{f16}
\end{figure}

\begin{figure}
\epsscale{1.0} \figurenum{17} \plotone{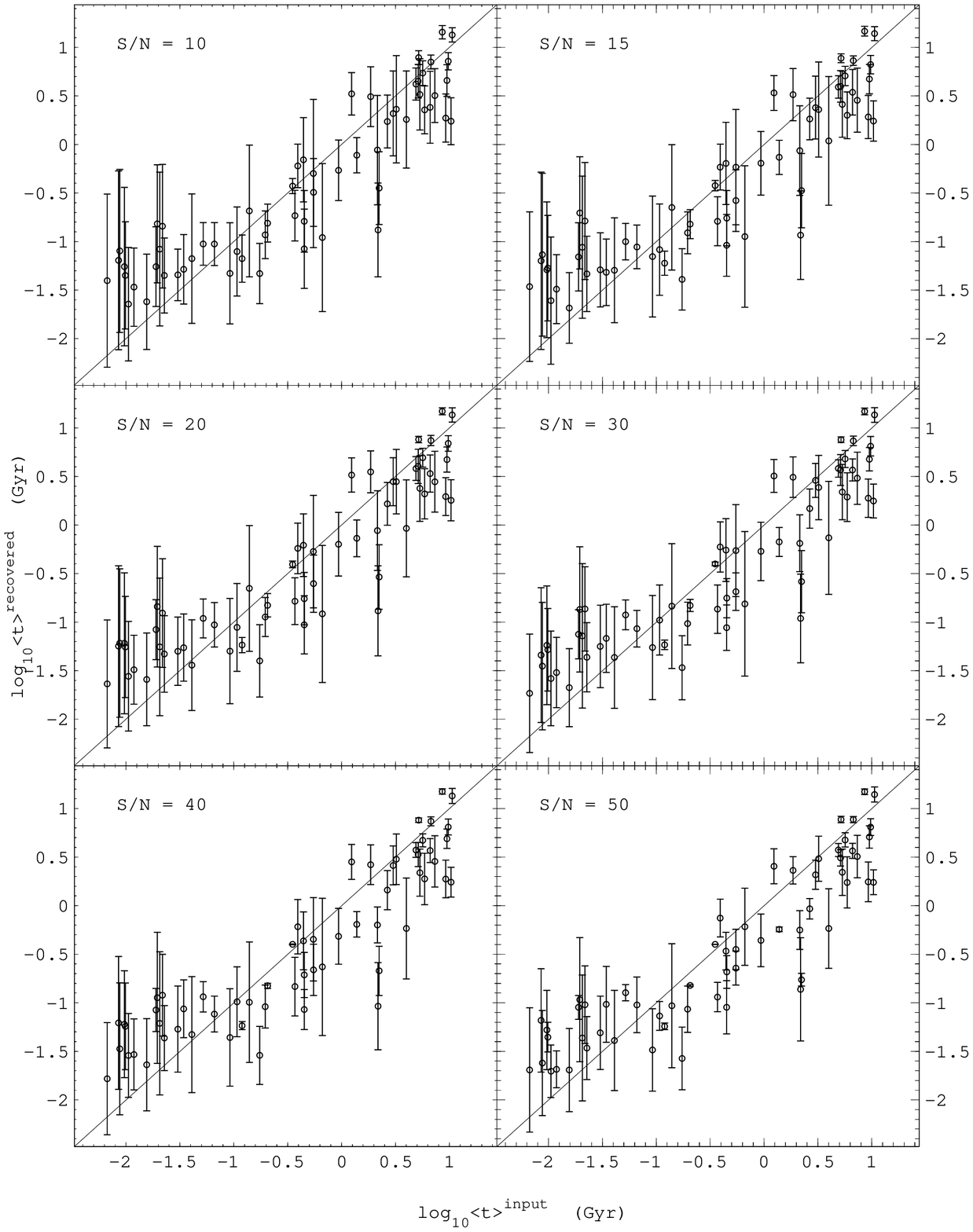}
\caption{Comparison of the recovered and input light-weighted
average age.  The error bar denotes $1~\sigma$ rms of the
recovered light-weighted average age. } \label{f17}
\end{figure}

\begin{figure}
\epsscale{1.0} \figurenum{18} \plotone{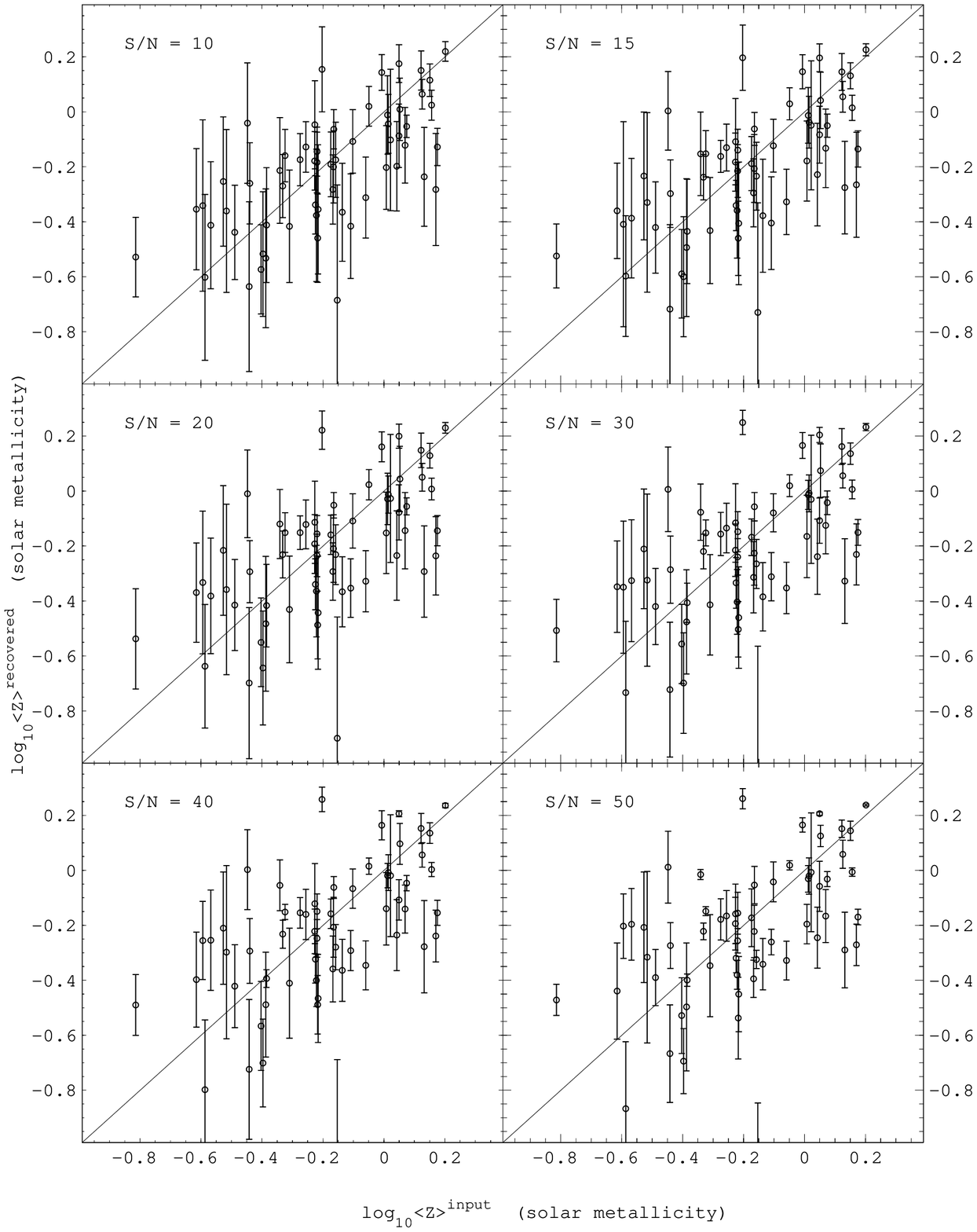}
\caption{Comparison of the recovered and input light-weighted
metallicity. } \label{f18}
\end{figure}

\begin{figure}
\epsscale{1.0} \figurenum{19} \plotone{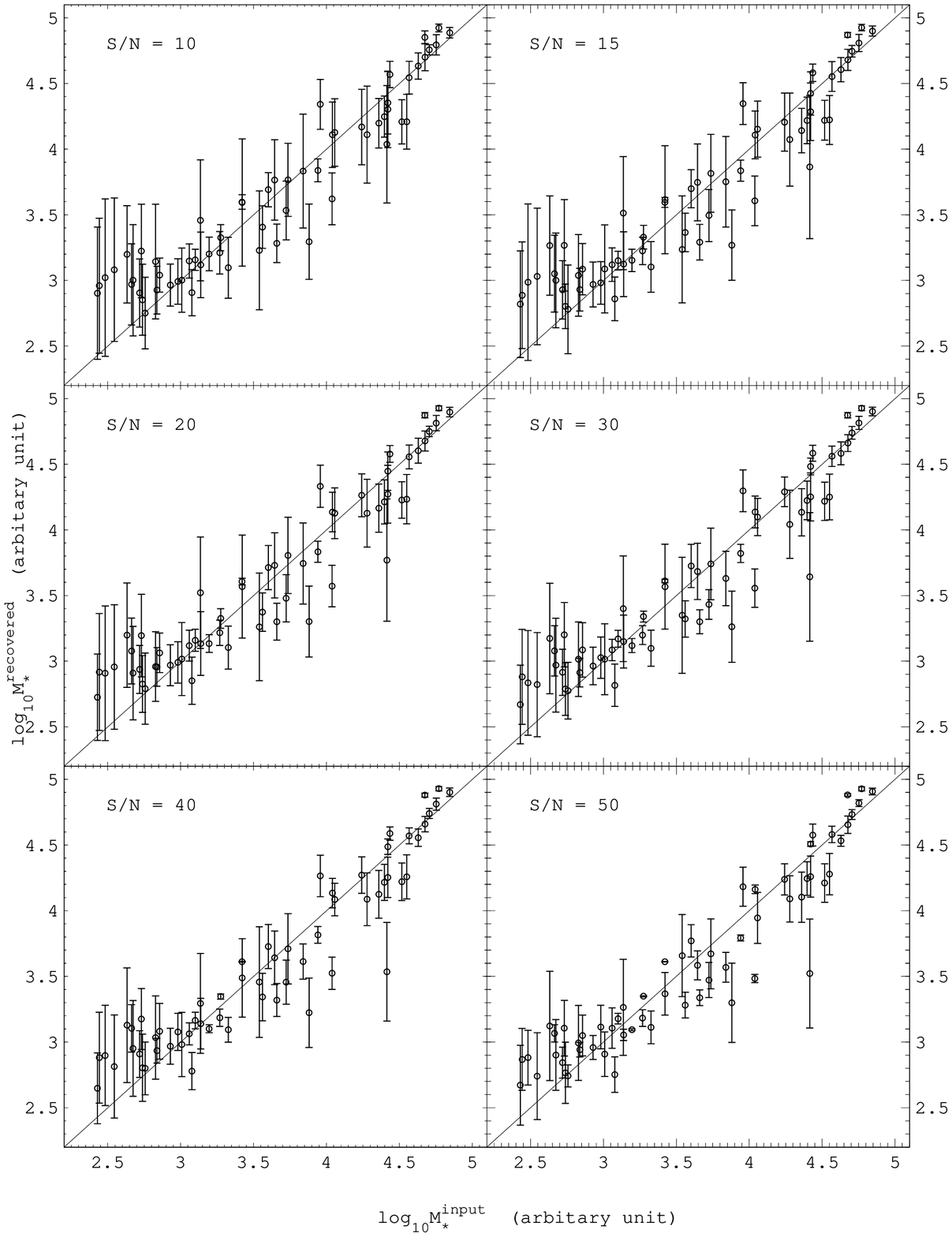}
\caption{Comparison of the recovered and input stellar mass.  }
\label{f19}
\end{figure}

\begin{figure}
\epsscale{1.0} \figurenum{20} \plotone{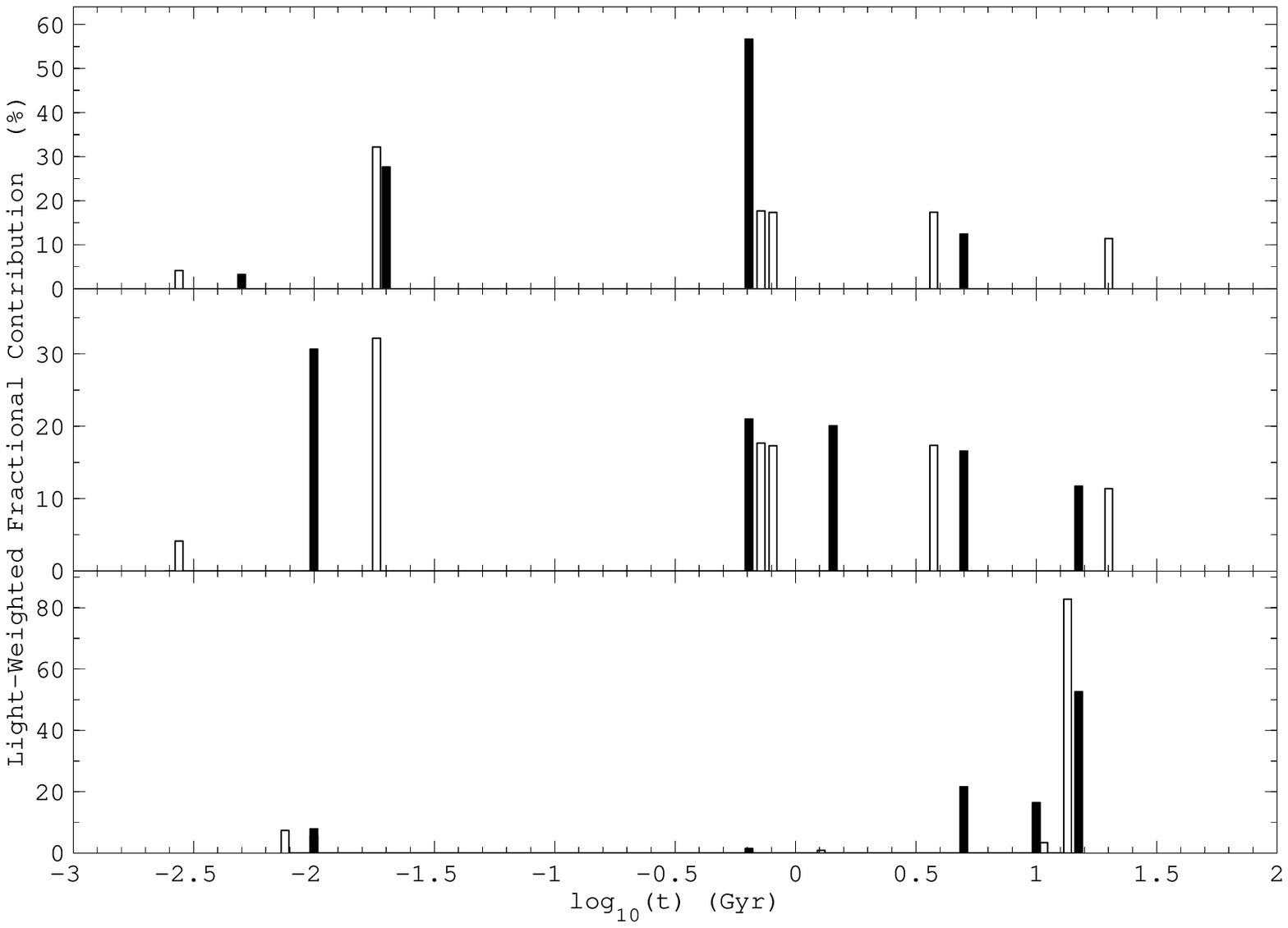}
\caption{Comparison of the recovered (hollow) and input (filled)
star formation histories of 3 randomly selected synthesized
galaxies. } \label{f20}
\end{figure}

\clearpage

\begin{figure}
\epsscale{1.0} \figurenum{21} \plotone{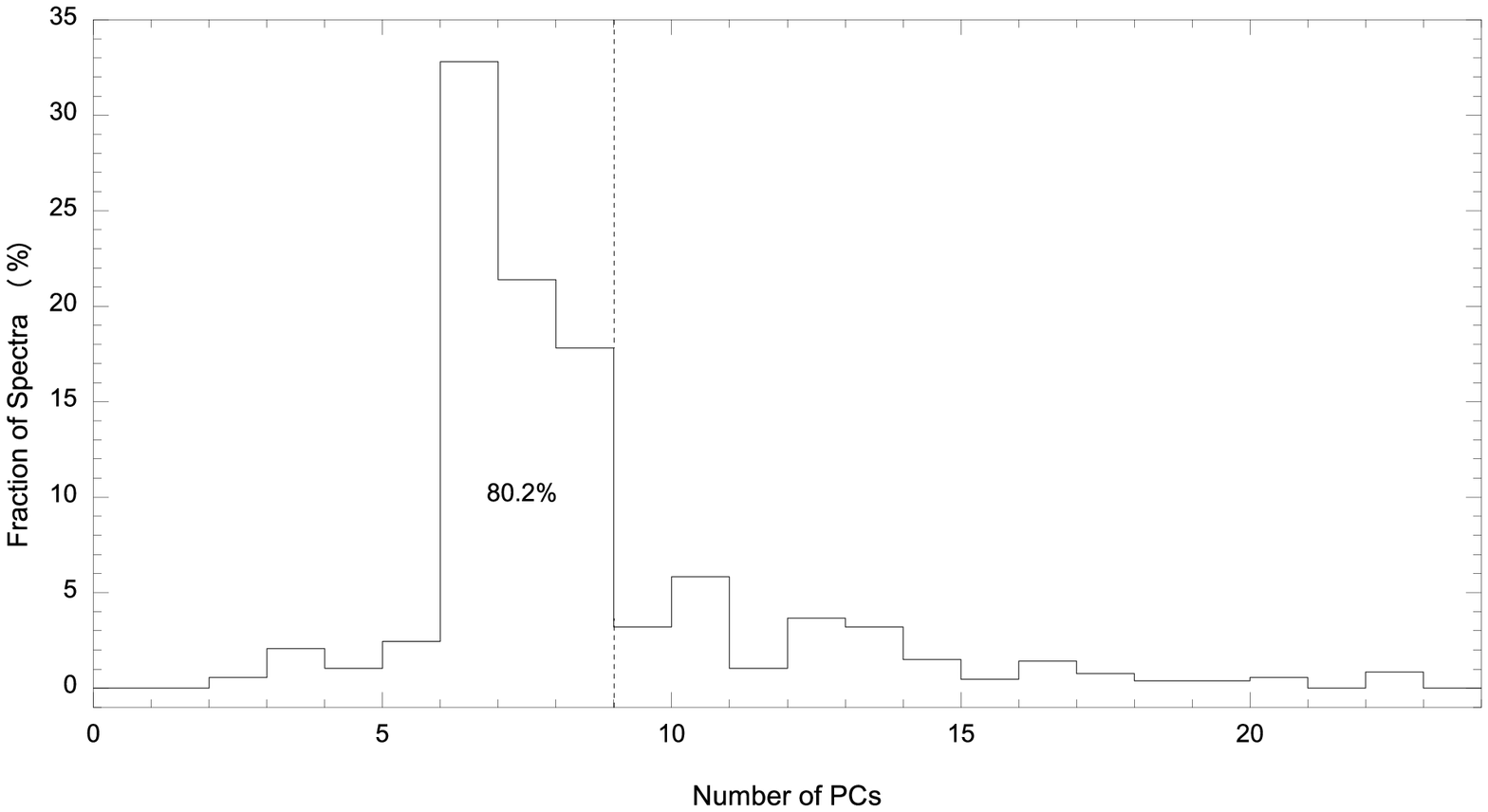}
\caption{Histogram of the F-test to determine the number of PCs
enough to represent GALXEV. } \label{f21}
\end{figure}

\begin{figure}
\epsscale{1.0} \figurenum{22} \plotone{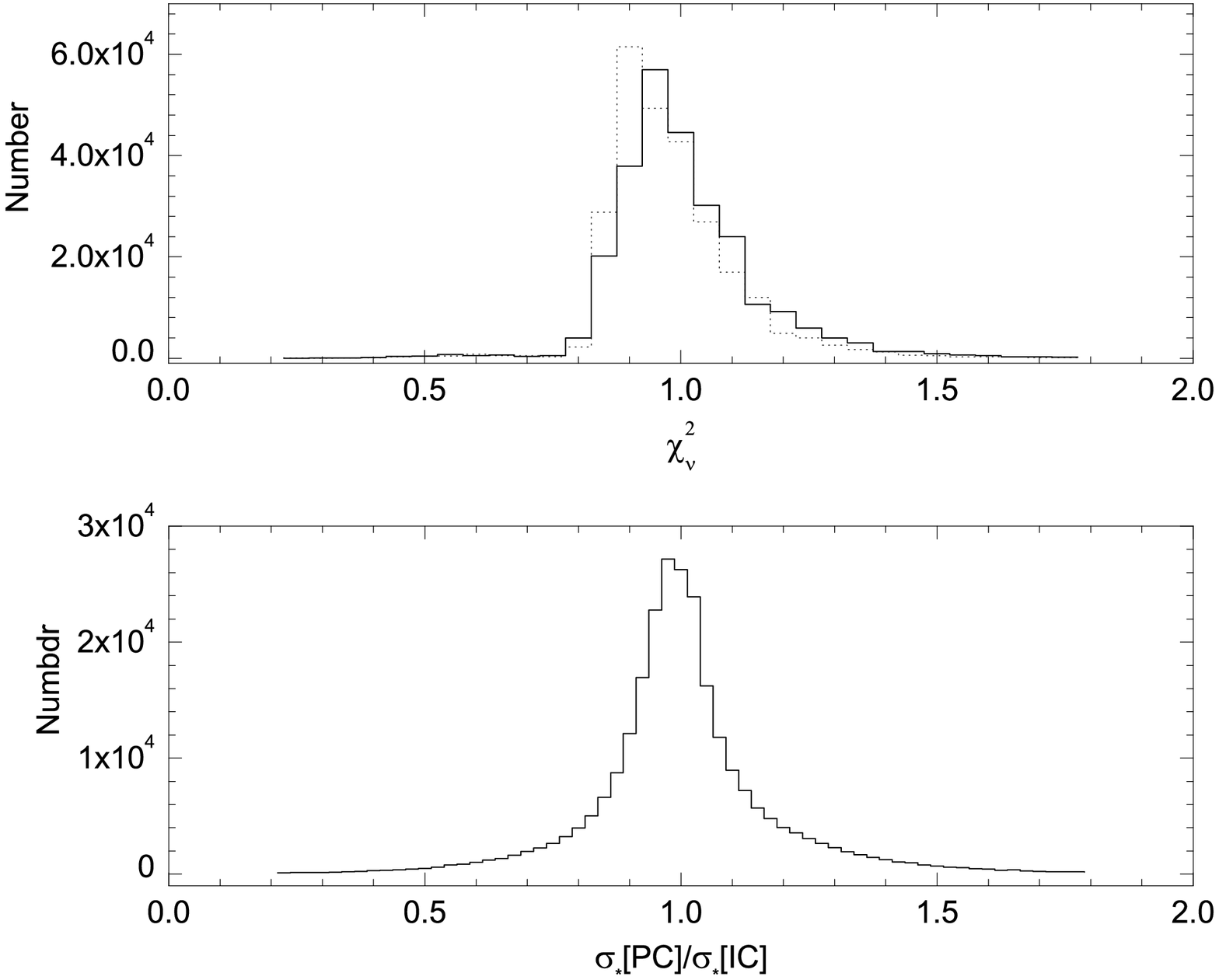} \caption{The
upper panel shows the distribution of reduced $\chi^2$ for PCA
model fits of all the $\sim 2.6\times 10^{5}$ galaxies in the SDSS
DR2 (solid line) compared with EL-ICA model (dotted line). Stellar
velocity dispersion obtained using the PCA model $\sigma_{*}[PC]$
and EL-ICA model $\sigma_{*}[IC]$ is compared in the lower panel.
} \label{f22}
\end{figure}

\begin{figure}
\epsscale{1.0} \figurenum{23} \plotone{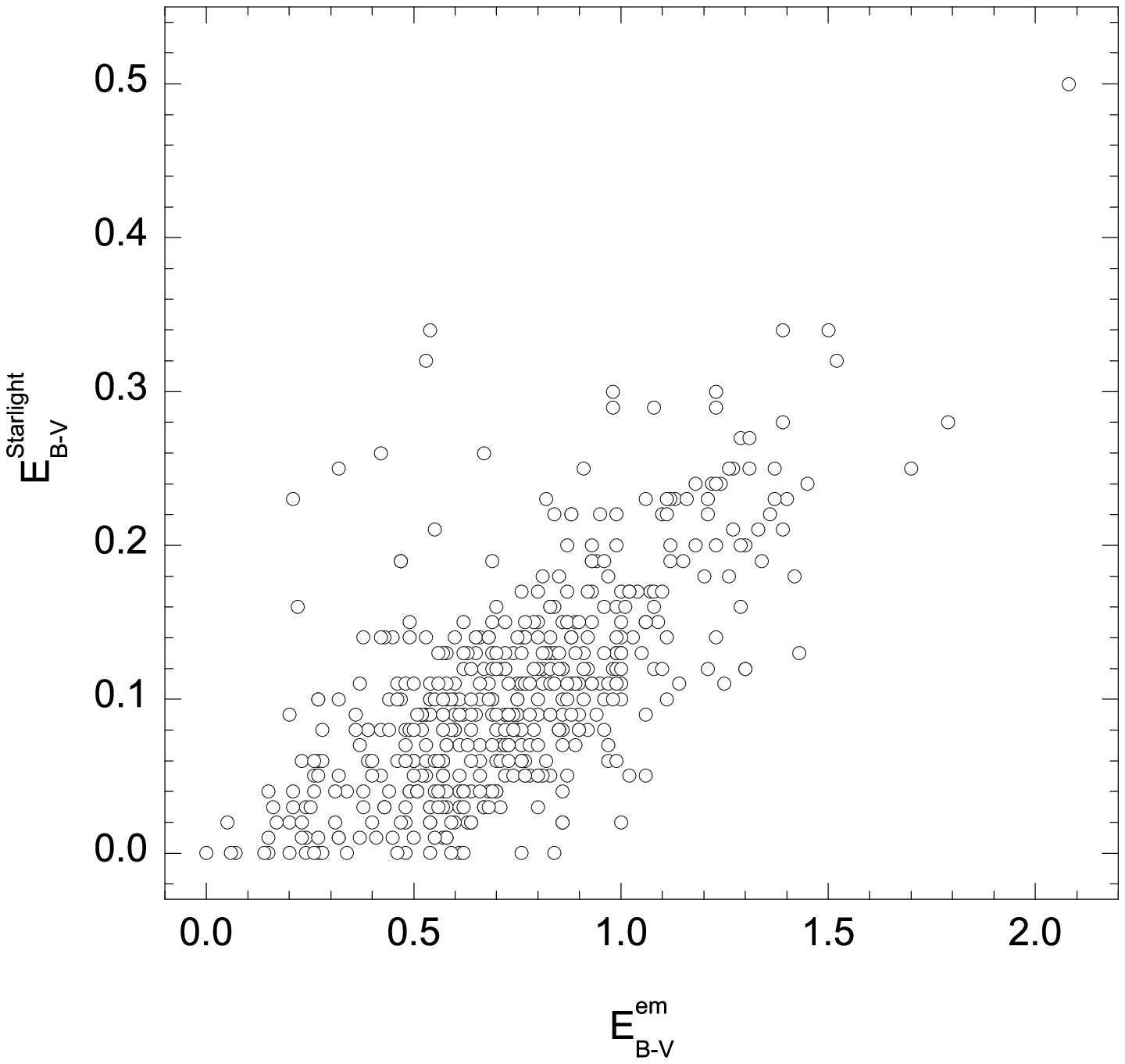} \caption{Same
as Figure \ref{f12} but the starlight reddening is derived using
PCA method. } \label{f23}
\end{figure}

\end{document}